\documentclass[aps,showpacs,floatfix,twocolumn,amsmath,amssymb]{revtex4}
\usepackage{epsfig}  

\begin{document}

\title{Octet baryon magnetic moments from QCD sum rules}
\author{Lai Wang and Frank X. Lee}
\affiliation{Physics Department, The George Washington University,
Washington, DC 20052, USA}

\begin{abstract}
A comprehensive study is made for the magnetic moments of octet baryons in
the method of QCD sum rules.
A complete set of QCD sum rules is derived using
the external field method and generalized interpolating fields.
For each member, three sum rules are constructed from three independent
tensor structures. They are analyzed in conjunction with the corresponding
mass sum rules. The performance of each of the sum rules is examined
using the criteria of OPE convergence and ground-state dominance, along with
the role of the transitions in intermediate states.
Individual contributions from the u, d and s quarks are isolated and
their implications in the underlying dynamics are explored.
Valid sum rules are identified and their predictions are obtained.
The results are compared with experiment and previous calculations.

\end{abstract}
\vspace{1cm}  
\maketitle

\section{Introduction}
\label{intro}
The QCD sum rule method is a nonperturbative analytic formalism firmly
entrenched in QCD with minimal modeling.
The field remains active judging by the
3000 and growing references to the seminal paper of
Shifman, Vainshtein and Zakharov~\cite{SVZ79} that introduced the method.
The approach provides a general way of linking
hadron phenomenology with the interactions of quarks and gluons via
only a few parameters: the QCD vacuum condensates and susceptibilities.
The studies give an unique perspective on how the properties of hadrons arise from
nonperturbative interactions in the QCD vacuum and how QCD works in this context.
It has been successfully applied in almost every aspect of strong-interaction physics.

Calculations of the magnetic moment were carried out soon after
 the method was introduced for the proton,
neutron~\cite{Balitsky83,Ioffe84} and hyperons~\cite{Ioffe83} in the
external field method. In this method, a static magnetic field is
introduced that couples to the quarks and polarizes the QCD
vacuum. Magnetic moments can be extracted from the linear response
to this field. The results of the studies validated the external
field method as a way of probing hadron properties other than the
mass, such as magnetic moments, form factors, axial charge,
isospin-breakings. Later, a more systematic study was made for the
magnetic moments of octet
baryons~\cite{Chiu86,Pasupathy86,Wilson87,SYZ}. Calculations were
also carried out for decuplet baryons~\cite{Lee98b,Lee98c,JMA,MAM}
and the rho meson~\cite{AS2002}.  There are other studies of
magnetic moments using the light-cone QCD sum rule
method~\cite{TIM,TAM02,TAM01,TAM00} which will not discuss here.

In this work, we carry out a comprehensive, independent calculation
of the magnetic moments of the octet baryons in the external field
method. It can be considered as an update over the previous
calculations~\cite{Chiu86,Pasupathy86,Wilson87} which were done more
than 20 years ago. There are a number of things we do differently.
First, we employ generalized interpolating fields which allow us to
use the optimal mixing of interpolating fields to achieve the best
match. Second, we derive a new, complete set of QCD sum rules at all
three tensor structures and analyze all of them. The previous sum
rules, which were mostly limited to one of the tensor structures,
correspond to a special case of the mixing in our sum rules. In this
way, we provide an independent check of the previous sum rules.
Third, we perform a Monte-Carlo analysis which has become standard
nowadays. The advantage of such an analysis is explained later.
Fourth, we use a different procedure to extract the magnetic moments
and to treat the transition terms in the intermediate states. Our
results show that these transitions cannot be simply ignored. Fifth,
we isolate the individual quark contributions to the magnetic
moments and discuss their implications in the underlying
quark-gluon dynamics in the baryons.

The paper is organized as follows. In Section~\ref{meth},
the method of QCD sum rules is introduced. Using the
interpolating fields for the octet baryons, master formula are
calculated. Then both the phenomenological representation and QCD
side are derived. Section~\ref{qcdsr} will list the sum rules we derived for the
octet baryon family, followed by the analysis to extract the magnetic moments
in Section~\ref{ana}. Section~\ref{res} summarizes the results
and gives a comparison of our results with experiment and previous calculation,
followed by an in-depth discussion of our findings. Our conclusions are given
in Section~\ref{con}.

\section{Method}
\label{meth}

The starting point is the time-ordered
correlation function in the QCD vacuum in the presence of a {\em
constant} background electromagnetic field $F_{\mu\nu}$:
\begin{equation}
\Pi(p)=i\int d^4x\; e^{ip\cdot x} \langle 0\,|\,
T\{\;\eta(x)\, \bar{\eta}(0)\;\}\,|\,0\rangle_F.
\label{cf2pt}
\end{equation}
The QCD sum rule approach is to evaluate this correlation at
two different levels. On the quark level, it describes a hadron
as quarks and gluons interacting in the QCD vacuum. On the phenomenological level,
It is saturated by a tower of hadronic intermediate states with the same quantum numbers.
This way, a connection can be established between a description in
terms of hadronic degrees of freedom and one based on the underlying
quark and gluon degrees of freedom governed by QCD.
Here $\eta$ is the interpolating field (or hadron current) with the
quantum numbers of the hadron under consideration. The subscript $F$
means that the correlation function is to be evaluated with an
electromagnetic interaction term added to the QCD Lagrangian:
\begin{equation}
{\cal L}_I = - A_\mu J^\mu,
\end{equation}
where $A_\mu$ is the external electromagnetic potential and
$J^\mu=e_q \bar{q} \gamma^\mu q$ is the quark electromagnetic current.

Since the external field can be made arbitrarily small, one can
expand the correlation function
\begin{equation}
\Pi(p)=\Pi^{(0)}(p) +\Pi^{(1)}(p)+\cdots,
\end{equation}
where $\Pi^{(0)}(p)$ is the correlation function in
the absence of the field, and gives rise to the mass sum rules of
the baryons. The magnetic moments will be extracted from the QCD sum
rules obtained from the linear response function
$\Pi^{(1)}(p)$.

The action of the external electromagnetic field is two-fold: it
couples directly to the quarks in the baryon interpolating fields,
and it also polarizes the QCD vacuum. The latter can be described
by introducing new parameters called  vacuum susceptibilities.

The interpolating field is constructed from quark fields with the
quantum number of baryon under consideration and it is not unique.
We consider a linear combination of the two standard local interpolating fields.
They read for the baryon octet family:
\begin{equation}
\begin{array}{l}
\eta^{p}(uud)= -2\epsilon^{abc} [(u^{aT}C\gamma_5
d^b)u^c+\beta (u^{aT}C d^b)\gamma_5 u^c],
\\
\eta^{n}(ddu)= -2\epsilon^{abc} [(d^{aT}C\gamma_5
u^b)d^c+\beta (d^{aT}C u^b)\gamma_5 d^c],
\\
\eta^{\Lambda}(uds)={-2\sqrt{\frac{1}{6}}}\epsilon^{abc} [
2(u^{aT}C\gamma_5 d^b)s^c +(u^{aT}C\gamma_5 s^b)d^c
\\
-(d^{aT}C\gamma_5 s^b)u^c \hspace{2mm}+\beta(2(u^{aT}C
d^b)\gamma_5s^c +(u^{aT}C s^b)\gamma_5d^c \\ -(d^{aT}C
s^b)\gamma_5u^c )],
\\
\eta^{\Sigma^-}(dds)=-2\epsilon^{abc} [(d^{aT}C\gamma_5
s^b)d^c+\beta (d^{aT}C s^b)\gamma_5 d^c],
\\
\eta^{\Sigma^0}(uds)=-\sqrt{2}\epsilon^{abc} [
 (u^{aT}C\gamma_5 s^b)d^c
+(d^{aT}C\gamma_5 s^b)u^c \\ +\beta((u^{aT}C s^b)\gamma_5d^c +(d^{aT}C
s^b)\gamma_5u^c) ],
\\
\eta^{\Sigma^+}(uus)=-2\epsilon^{abc} [(u^{aT}C\gamma_5
s^b)u^c+\beta  (u^{aT}C s^b)\gamma_5 u^c],
\\
\eta^{\Xi^-}(ssd)=-2\epsilon^{abc} [(s^{aT}C\gamma_5
d^b)s^c+\beta  (s^{aT}C d^b)\gamma_5 s^c],
\\
\eta^{\Xi^0}(ssu)=-2\epsilon^{abc} [(s^{aT}C\gamma_5
u^b)s^c+\beta  (s^{aT}C u^b)\gamma_5 s^c].
\end{array}
\label{infield}
\end{equation}
Here $u$ and $d$ are up-quark and down-quark field operators, $C$ is
the charge conjugation operator, the superscript $T$ means
transpose, and $\epsilon_{abc}$ makes it color-singlet. The
normalization factors are chosen so that correlation functions of
these interpolating fields coincide with each other under
SU(3)-flavor symmetry. The real parameter $\beta$ allows for the
mixture of the two independent currents. The choice advocated by
Ioffe \cite{Ioffe81} and often used in QCD sum rules studies corresponds to $\beta=-1$.
We will take advantage of this freedom to achieve optimal matching in the sum rule analysis.

\begin{widetext}
\subsection{Phenomenological Representation}
\label{rhs}

We start with the structure of the two-point
correlation function in the presence of the electromagnetic vertex
to first order
\begin {equation}
\Pi  (p) = i\int {d^4 x} e^{ipx}  < 0|\eta  (x)[ - i\int {d^4 y}
A_\mu  (y)J^\mu  (y)]\bar \eta(0)|0>.
\end {equation}
Inserting two complete sets of physical intermediate states, we
restrict our attention only to the positive energy ones and write
\begin {equation}
\begin{array}{l}
 \Pi(p) = \int {d^4 x} d^4 y\frac{{d^4 k'}}{{(2\pi )^4 }}\frac{{d^4 k}}{{(2\pi )^4 }}\sum\limits_{N'N} {\sum\limits_{s's} {\frac{{ - i}}{{k'^2  - M_{N'} ^2  - i\varepsilon }}} } \frac{{ - i}}{{k^2  - M_{N}^2  - i\varepsilon }} \\
 e^{ipx} A_{\mu} (y) < 0|\eta   (x)|N'k's' >  < N'k's'|J^\mu  (y)| Nks>  <Nks|\bar \eta  (0)|0 > . \\
\end{array}
\end {equation}
We can use the translation invariance to express $\eta (x)$ in terms of $\eta (0)$
\begin {equation}
\begin{array}{l}
< 0|\eta   (x)|N'k's' >=< 0|\eta (0)|N'k's' >e^{-ik'x}.
\end{array}
\end {equation}
The interpolating field excites (or annihilates) the ground state as
well as the excited states of the baryon from the QCD vacuum. The
ability to do so is described by a phenomenological parameter $\lambda_N$
(called current coupling or pole residue),
defined by the overlap for the ground state
\begin {equation}
\begin{array}{l}
 < 0|\eta   (0)|Nks >=\lambda _{N} u (k,s),
\end{array}
\end {equation}
where $u$ is the Dirac spinor.

Translation invariance on $J^\mu  (y)$ gives
\begin {equation}
\begin{array}{l}
< N'k's'|J^\mu  (y)|Nks >=e^{iqy}< N'k's'|J^\mu (0)|Nks>,
\end{array}
\end {equation}
where $q=k'-k$ is the momentum transfer and $Q^2=-q^2$.

The matrix element of the electromagnetic current has the general form
\begin {equation}
\begin{array}{l}
< k's'|J^\mu  (0)| ks>=\bar
u(k',s')[F_1(Q^2)\gamma_\mu+F_2(Q^2)i\sigma^{\mu\nu}\frac{q^\nu}{2M_N}]u(k,s),
 \end{array}
\end {equation}
where the Dirac form factors $F_1$ and $F_2$ are related to the Sachs form factors by
\begin {equation}
\begin{array}{l}
G_E(Q^2)=F_1(Q^2)-\frac{Q^2}{(2M_N)^2}F_2(Q^2)\\
G_M(Q^2)=F_1(Q^2)+F_2(Q^2).
 \end{array}
\end {equation}
At $Q^2=0$, $F_1(0)=1$, $F_2(0)=\mu^a$ which is the anomalous magnetic moment,
and $G_M(0)=F_1(0)+F_2(0)=\mu$ which is the total magnetic moment.

Writing out explicitly only the contribution of the ground-state
nucleon and denoting the excited state contribution by ESC, we have
\begin {equation}
\begin{array}{l}
 \Pi  (p) =  - \lambda _N ^2 \int {d^4 x} d^4 y\frac{{d^4 k'}}{{(2\pi )^4 }}\frac{{d^4 k}}{{(2\pi )^4 }}[k'^2  - M_N ^2  - i\varepsilon ]^{ - 1} [k^2  - M_{N}^2  - i\varepsilon ]^{ - 1}
 A_\mu  (y)e^{i(p - k')x} \\e^{iqy}
 \sum\limits_{s'}  {u    (k',s' )\bar u  (k',s' )[F_1(Q^2)\gamma^\mu+F_2(Q^2)i\sigma^{\mu\nu}\frac{q^\nu}{2M_N}] } \sum\limits_{s } {u(k,s)\bar u  (k,s )}  + ESC. \\
 \end{array}
\end {equation}
The spin sums are of the form
\begin {equation}
\begin{array}{l}
 \sum\limits_{s} {u(k,s)\bar u(k,s  )}  = \hat k + M_N.
 \end{array}
\end {equation}
QCD sum rule calculations are most conveniently done in the
fixed-point gauge. For electromagnetic field, it is defined by
$x_\mu  A_\mu (x) = 0$. In this gauge, the electromagnetic potential
is given by
\begin {equation}
A_\mu  (y) =  - \frac{1}{2}F_{\mu \nu } y^\nu.
\end {equation}
Changing variables from $k$ to $q=k'-k$, then $d^4 k =- d^4 q$, we have
\begin {equation}
\begin{array}{l}
 \Pi (p) = \frac{- \lambda _N^2 }{2}F_{\mu \nu } \int {d^4 x} d^4 y\frac{{d^4 k'}}{{(2\pi )^4 }}\frac{{d^4 q}}{{(2\pi )^4 }}[k'^2  - M_N ^2  - i\varepsilon ]^{-1}[(q - k')^2  - M_N^2  - i\varepsilon ]^{ - 1}
 \\\times e^{i(p - k')x} ( - i\frac{\partial }{{\partial q_\nu  }}e^{iq\cdot y} )(\hat p + M_N  )[F_1(Q^2)\gamma_\mu+F_2(Q^2)i\sigma^{\mu\nu}\frac{q^\nu}{2M_N}] (\hat {k'} - \hat q + M_N ) + ESC. \\
 \end{array}
\end {equation}
Integrating over x, we get a delta function
\begin {equation}
\int {d^4 x} \frac{1}{{(2\pi )^4 }}e^{i(p - k')x} = \delta ^4 (p -
k').
\end {equation}
Integrating $\partial/\partial q_\nu $ by parts, then doing $\int
{d^4 y}$, we can get another delta function $\delta (q)$. Since we
have a $\delta (q)$, when doing $\partial/\partial q_\nu $ only
terms linear in $q_\nu $ contribute. We have $F_1 (Q^2)|_{q=0}=1,F_2
(Q^2)|_{q=0}=\mu^a=\mu-1$, and $\partial/\partial q_\nu F_{1,2} (Q^2)|_{q=0}=0$,
so derivatives of the structure functions do not enter.
Finally, we arrive at
\begin {equation}
\begin{array}{l}
 \Pi (p) = \frac{{i}}{2}\lambda _N^2 F_{\mu \nu } [p^2  - M_N^2  - i\varepsilon ]^{ - 2} (\hat p + M_N )\{ \frac{{i(\mu_N-1)}}{{2M_N }}\sigma ^{\mu \nu } (\hat p + M_N ) + i\sigma ^{\mu \nu }  \\
  -( p^\mu  \gamma ^\nu   - p^\nu  \gamma ^\mu  )(\hat p + M_N )[p^2  - M_N^2  - i\varepsilon ]^{ - 1} \}  +
  ESC.
 \end{array}
\end {equation}
Examination of its tensor structure reveals that its has 3
independent combinations: $ F^{\mu \nu } (\hat p\sigma _{\mu \nu } +
\sigma _{\mu \nu } \hat p)$,  $F^{\mu \nu }i (p_\mu \gamma _\nu -
p_\nu \gamma _\mu )\hat p$ and  $F^{\mu \nu } \sigma _{\mu \nu }$.
The momentum-space correlation function in the above equation
can be written in terms of these three structures
\begin {equation}
\begin{array}{l}
 \Pi (p) =  - \frac{1}{4}\frac{{\lambda _N^2 F_{\mu \nu } }}{{[p^2  - M_N^2  - i\varepsilon ]^2 }}\{ [\sigma ^{\mu \nu } ](2M_N \mu _N ) + \frac{{\mu _N  - 1}}{{M_N }}(p^2  - M_N^2 ) + \mu _N [\hat p\sigma ^{\mu \nu }  + \sigma ^{\mu \nu } \hat p] \\
  + \frac{{2(\mu _N  - 1)}}{{M_N }}[i(p^\mu  \gamma ^\nu   - p^\nu  \gamma ^\mu  )]\hat p\}  + ESC, \\
 \end{array}
\end {equation}
where we have used the following identities
\begin {equation}
\begin{array}{l}
 (\hat p + M_N )\sigma ^{\mu \nu }  = M_N \sigma ^{\mu \nu }  + \frac{1}{2}(\hat p\sigma ^{\mu \nu }  + \sigma ^{\mu \nu } \hat p) + i(p^\mu  \gamma ^\nu   - p^\nu  \gamma ^\mu  ), \\
 (\hat p + M_N )\sigma ^{\mu \nu } (\hat p + M_N ) = \sigma ^{\mu \nu } (p^2  + M_N^2 ) + M_N (\hat p\sigma ^{\mu \nu }  + \sigma ^{\mu \nu } \hat p) + 2i(p^\mu  \gamma ^\nu   - p^\nu  \gamma ^\mu  )\hat p, \\
(\hat p + M_N )(p^\mu  \gamma ^\nu   - p^\nu  \gamma ^\mu  )(\hat p + M_N ) =  - (p^2  - M_N^2 )(p^\mu  \gamma ^\nu   - p^\nu  \gamma ^\mu  ). \\
  \end{array}
\end {equation}
Next step is to perform the Borel transform defined  by
\begin{equation}
\hat{B}[f(p^2)] = \mathop{\lim}_{\scriptstyle {-p^2, n \rightarrow \infty}\hfill\atop \scriptstyle {-p^2/n=M^2}\hfill}
\frac{1}{n!} (-p^2)^{n+1}(\frac{d}{dp^2})^n f(p^2).
\label{borel_trans}
\end{equation}
%
Upon Borel transform the ground state takes the form
\begin {equation}
\begin{array}{l}
 \hat B[\Pi (p)] =  - \frac{{\lambda _{\rm N}^2 }}{{4M^2 }}e ^{ - M_N^2 /M^2 } \{ \frac{1}{{M_N }}(2M_N^2 \mu_N - M^2 (\mu_N-1) )[F_{\mu \nu } \sigma ^{\mu \nu } ] +  \\
 \mu_N[F_{\mu \nu } (\hat p\sigma ^{\mu \nu }  + \sigma ^{\mu \nu } \hat p)] + \frac{{2(\mu_N-1)}}{{M_N }}[F_{\mu \nu } i(p^\mu  \gamma ^\nu   - p^\nu  \gamma ^\mu  )\hat
 p]\},
  \end{array}
\end {equation}
where $M$ is the Borel mass, not to be confused with the nucleon mass $M_N$.

Here we must treat the excited states with care. For a generic
invariant function, the pole structure can be written as
\begin {equation}
\begin{array}{l}
  \frac{C_{N \leftrightarrow N }^2}{(p^2  -
M_N^2 )^2}+ \sum\limits_{N^* } {\frac{{C_{N \leftrightarrow N^* }^2
}}{{(p^2  - M_N^2 )(p^2  - M_{N^* }^2 )}}}  + \sum\limits_{N^* }
{\frac{{C_{N^* \leftrightarrow N^* }^2 }}{{(p^2  - M_{N^* }^2 )^2 }}},
\end{array}
\end {equation}
 where $C_{N \leftrightarrow N}$, $C_{N \leftrightarrow N^*}$ and $C_{N^* \leftrightarrow N^*}$ are constants. The first term is
 the ground state pole which contains the desired magnetic
 moment $\mu_N$. The second term represents the non-diagonal
 transitions between the ground state and the excited states caused
 by the external field. The third term is pure excited state contributions.
These different contributions can be represented by the diagrams in Fig.~\ref{3states}.
 %
\begin{figure}
\centerline{\psfig{file=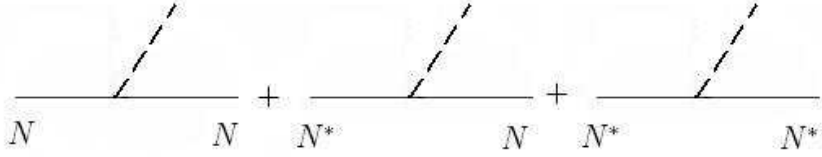,width=12.0cm}} \vspace*{-0.5cm}
\caption{\small{The three kinds of contributions to the spectral
function in the presence of an external field:
ground state, transitions between ground state and excited states, and
pure excited states}} \label{3states}
\end{figure}
%
Upon Borel transform, it takes the form
\begin {equation}
\begin{array}{l}
\frac{\lambda _N^2\mu
_N}{M^2}e^{-M_N^2/M^2}+e^{-M_N^2/M^2}\left[\sum\limits_{N^* }
\frac{C_{N\leftrightarrow
N^*}^2}{M_{N^*}^2-M_N^2}\left(1-e^{-\left(M_{N^*}^2-M_N^2\right)/M^2}\right)\right]+\sum
\limits_{N^* } \frac{C_{N^*\leftrightarrow
N^*}^2}{M^2}e^{-M_{N^*}^2/M^2}.
\end{array}
\end {equation}
The important point is that the transitions give rise to a
contribution that is not
 exponentially suppressed relative to the ground state. This is a
 general feature of the external-field technique. The strength of
 such transitions at each structure is a \emph{priori} unknown and is an
 additional source of contamination in the determination of the magnetic moment
 $\mu_N$. The standard treatment of the transitions is to approximate the quantity in the square brackets
 by a constant, which is to be extracted from the sum rule along
 with the ground state property of interest. Inclusion of such
 contributions is necessary for the correct extraction of the
 magnetic moments. The pure excited state contributions are
 exponentially suppressed relative to the ground state and can be
 modeled in the usual way by introducing a continuum model and
 threshold parameter.


\subsection{Calculation of the QCD Side}
\label{lhs}

We start by contracting out the quark pairs in Eq.~(\ref{cf2pt}) using
Wick's theorem, resulting in the so-called {\em master formula} in terms of
quark propagators.
The master formula for the proton (with $uud$ quark content) is
\begin{eqnarray}
& & \langle\Omega\,|\, T\{\eta^{N}(x)\,
\bar{\eta}^{N}(0)\,|\,\Omega\rangle
=-4\epsilon^{abc}\epsilon^{a^\prime b^\prime c^\prime} \{\;
\nonumber \\ & & +S^{aa^\prime}_u \gamma_5 C {S^{cc^\prime}_d}^T C
\gamma_5 S^{bb^\prime}_u +S^{aa^\prime}_u \mbox{Tr}(C
{S^{cc^\prime}_d}^T C \gamma_5 S^{bb^\prime}_u \gamma_5) \nonumber
\\ & & +\beta\gamma_5  S^{aa^\prime}_u \gamma_5 C
{S^{cc^\prime}_d}^T C S^{bb^\prime}_u +\beta \gamma_5
S^{aa^\prime}_u \mbox{Tr}( C {S^{cc^\prime}_u}^T C S^{bb^\prime}_d
\gamma_5) \nonumber \\ & & +\beta  S^{aa^\prime}_u C
{S^{cc^\prime}_d}^T C \gamma_5 S^{bb^\prime}_u \gamma_5 +\beta
S^{aa^\prime}_u \gamma_5 \mbox{Tr}( C {S^{cc^\prime}_u}^T C \gamma_5
S^{bb^\prime}_d) \nonumber \\ & & +\beta^2 \gamma_5 S^{aa^\prime}_u
C {S^{cc^\prime}_d}^T C S^{bb^\prime}_u \gamma_5 +\beta^2 \gamma_5
S^{aa^\prime}_u \gamma_5 \mbox{Tr}( C {S^{cc^\prime}_d}^T C
S^{bb^\prime}_u) \;\}. \label{master11}
\end{eqnarray}
The master formula for neutron (with $ddu$ quark content) can be
obtained by exchanging the $d$ quark with a $u$ quark from
Eq.~(\ref{masterll}). By replacing the $d$ quark with a $s$
quark, one can get the master formula for $\Sigma^+$ (with $uus$
quark content). While the master formula for $\Sigma^-$ (with $uus$
quark content) can be obtained by replacing the $u$ quark with a $d$
quark from $\Sigma^+$ master formula. By exchanging the $u$ quarks
and $s$ quarks in $\Sigma^+$ master formula, the master formula for
$\Xi^0$ (with $ssu$ quark content) can be obtained. Likewise, by
replacing the $u$ quark with a $d$ quark, one can get the master
formula for $\Xi^+$ (with $ssd$ quark content).

The master formulae for $\Sigma^0$ and $\Lambda$ ($uds$
quark content) have a more complicated structure. They can be
written in a combined way as
\begin{eqnarray}
& & \langle\Omega\,|\, T\{\;\eta(x)\, \bar{\eta}(0)\;\}
\,|\,\Omega\rangle =-f\epsilon^{abc}\epsilon^{a^\prime b^\prime c^\prime} \{\; \nonumber
\\ & &
 f_1 f_2 S^{aa^\prime}_s \gamma_5 C {S^{cc^\prime}_u}^T
C\gamma_5S^{bb^\prime}_d
  -f_1 f_3 S^{aa^\prime}_s \gamma_5 C{S^{cc^\prime}_d}^T C \gamma_5 S^{bb^\prime}_u
 +f_1 f_1 S^{aa^\prime}_s\mbox{Tr}(C {S^{cc^\prime}_u}^T C \gamma_5 S^{bb^\prime}_d \gamma_5)
\nonumber
\\ & &
 +f_2 f_1 S^{aa^\prime}_d \gamma_5 C {S^{cc^\prime}_u}^T C\gamma_5S^{bb^\prime}_s
 +f_2 f_3 S^{aa^\prime}_d \gamma_5 C{S^{cc^\prime}_s}^T C \gamma_5S^{bb^\prime}_u
 +f_2 f_2 S^{aa^\prime}_d \mbox{Tr}(C {S^{cc^\prime}_s}^T C \gamma_5 S^{bb^\prime}_u \gamma_5)
\nonumber
\\ & &
 -f_3 f_1 S^{aa^\prime}_u \gamma_5 C {S^{cc^\prime}_d}^T C  \gamma_5S^{bb^\prime}_s
 +f_3 f_2 S^{aa^\prime}_u \gamma_5 C {S^{cc^\prime}_s}^T C\gamma_5 S^{bb^\prime}_d
 +f_3 f_3 S^{aa^\prime}_u \mbox{Tr}(C{S^{cc^\prime}_s}^T C \gamma_5 S^{bb^\prime}_d \gamma_5)
 \nonumber
\\ & &
 +f_1 f_5 \beta S^{aa^\prime}_s C {S^{cc^\prime}_u}^T C  \gamma_5S^{bb^\prime}_d \gamma_5
 -f_1 f_6 \beta S^{aa^\prime}_s C{S^{cc^\prime}_d}^T C  \gamma_5 S^{bb^\prime}_u \gamma_5
 +f_1 f_4 \beta S^{aa^\prime}_s \gamma_5 \mbox{Tr}(C {S^{cc^\prime}_d}^T C \gamma_5 S^{bb^\prime}_u)
 \nonumber
 \\ & &
 +f_2 f_4 \beta S^{aa^\prime}_d C{S^{cc^\prime}_u}^T C \gamma_5 S^{bb^\prime}_s \gamma_5
 +f_2 f_6 \beta S^{aa^\prime}_d C {S^{cc^\prime}_s}^T C  \gamma_5 S^{bb^\prime}_u\gamma_5
 +f_2 f_5 \beta S^{aa^\prime}_d \gamma_5 \mbox{Tr}(C{S^{cc^\prime}_s}^T C \gamma_5 S^{bb^\prime}_u)
  \nonumber
  \\ & &
 -f_3 f_4 \beta S^{aa^\prime}_u C {S^{cc^\prime}_d}^T C  \gamma_5S^{bb^\prime}_s \gamma_5
 +f_3 f_5 \beta S^{aa^\prime}_u C{S^{cc^\prime}_s}^T C  \gamma_5 S^{bb^\prime}_d \gamma_5
 +f_3 f_6 \beta S^{aa^\prime}_u \gamma_5 \mbox{Tr}(C {S^{cc^\prime}_s}^T C \gamma_5 S^{bb^\prime}_d)
 \nonumber
 \\ & &
 +f_4 f_2 \beta \gamma_5 S^{aa^\prime}_s\gamma_5 C {S^{cc^\prime}_u}^T C S^{bb^\prime}_d
 -f_4 f_3 \beta \gamma_5 S^{aa^\prime}_s \gamma_5 C {S^{cc^\prime}_d}^T C S^{bb^\prime}_u
 +f_4 f_1 \beta \gamma_5 S^{aa^\prime}_s \mbox{Tr}(C {S^{cc^\prime}_d}^T C S^{bb^\prime}_u \gamma_5)
 \nonumber
 \\ & &
 +f_5 f_1 \beta \gamma_5 S^{aa^\prime}_d \gamma_5 C {S^{cc^\prime}_u}^T C S^{bb^\prime}_s
 +f_5 f_3 \beta \gamma_5 S^{aa^\prime}_d \gamma_5 C {S^{cc^\prime}_s}^T C S^{bb^\prime}_u
 +f_5 f_2 \beta \gamma_5 S^{aa^\prime}_d \mbox{Tr}(C{S^{cc^\prime}_s}^T C S^{bb^\prime}_u \gamma_5)
 \nonumber
 \\ & &
 -f_6 f_1 \beta \gamma_5 S^{aa^\prime}_u \gamma_5 C {S^{cc^\prime}_d}^T C S^{bb^\prime}_s
 +f_6 f_2 \beta \gamma_5 S^{aa^\prime}_u \gamma_5 C {S^{cc^\prime}_s}^T C S^{bb^\prime}_d
 +f_6 f_3 \beta \gamma_5S^{aa^\prime}_u \mbox{Tr}(C {S^{cc^\prime}_s}^T C S^{bb^\prime}_d\gamma_5)
 \nonumber
 \\ & &
 +f_4 f_5 \beta^2 \gamma_5 S^{aa^\prime}_s C{S^{cc^\prime}_u}^T C S^{bb^\prime}_d \gamma_5
 -f_4 f_6 \beta^2 \gamma_5 S^{aa^\prime}_s C {S^{cc^\prime}_d}^T C S^{bb^\prime}_u \gamma_5
 +f_4 f_4 \beta^2 \gamma_5 S^{aa^\prime}_s \gamma_5 \mbox{Tr}(C{S^{cc^\prime}_u}^T C S^{bb^\prime}_d)
 \nonumber
 \\ & &
 +f_5 f_4 \beta^2\gamma_5 S^{aa^\prime}_d C {S^{cc^\prime}_u}^T C  S^{bb^\prime}_s \gamma_5
 +f_5 f_6 \beta^2 \gamma_5 S^{aa^\prime}_d C {S^{cc^\prime}_s}^T C S^{bb^\prime}_u \gamma_5
 +f_5 f_5 \beta^2 \gamma_5 S^{aa^\prime}_d \gamma_5\mbox{Tr}(C {S^{cc^\prime}_s}^T C S^{bb^\prime}_u)
 \nonumber
 \\ & &
 -f_6 f_4 \beta^2 \gamma_5 S^{aa^\prime}_u C {S^{cc^\prime}_d}^T CS^{bb^\prime}_s \gamma_5
 +f_6 f_5 \beta^2 \gamma_5 S^{aa^\prime}_u C{S^{cc^\prime}_s}^T C S^{bb^\prime}_d \gamma_5
 +4 f_6 f_6 \beta^2 \gamma_5 S^{aa^\prime}_u \gamma_5 \mbox{Tr}(C {S^{cc^\prime}_s}^T C S^{bb^\prime}_d)\},
  \label{master11_lambda_o}
\end{eqnarray}
where the various factors are: for $\Sigma^0$,
$f=2/3, f_1=0, f_2=1, f_3=1, f_4=0, f_5=1, f_6=1$; and for $\Lambda$,
$f=2, f_1=2, f_2=1, f_3=-1, f_4=2, f_5=1, f_6=-1$.

In the above equations,
\begin{equation}
S^{ab}_q (x,0;F) \equiv \langle 0\,|\, T\{\;q^a(x)\,
\bar{q}^b(0)\;\}\,|\,0\rangle_F,
 \hspace{3mm} q=u, d, s,
\end{equation}
is the fully interacting quark propagator in the presence of the
electromagnetic field. To first order in $F_{\mu\nu}$ and $m_q$
(assume $m_u=m_d=0, m_s\neq 0$), and order $x^4$, it is given
by the operator product expansion (OPE)~\cite{Ioffe84,Pasupathy86,Wilson87}:
\begin{eqnarray}
S^{ab}_q(x,0;Z) &\equiv&
 {i \over 2\pi^2} {\hat{x}\over x^4} \delta^{ab}
- {m_q \over 4\pi^2 x^2} \delta^{ab} - {1\over
12}\langle\bar{q}q\rangle \delta^{ab} + {im_q \over 48}
\langle\bar{q}q\rangle \hat{x} \delta^{ab} \nonumber \\ & & +
{1\over 192} \langle\bar{q}g_c\sigma\cdot Gq\rangle x^2 \delta^{ab}
- {im_q\over 1152} \langle\bar{q}g_c\sigma\cdot Gq\rangle
    \hat{x} x^2 \delta^{ab}
- {1\over 3^3 2^{10}} \langle\bar{q}q\rangle \langle g^2_c
G^2\rangle
     x^4 \delta^{ab}
\nonumber \\ & & + {i\over 32\pi^2} (g_cG^n_{\alpha\beta})
  { \hat{x} \sigma^{\alpha\beta} +\sigma^{\alpha\beta} \hat{x}\over x^2 }
  \left({\lambda^n\over 2}\right)^{ab}
+ {1\over 48} {i\over 32\pi^2} \langle g^2_c G^2\rangle
  {\hat{x} \sigma^{\alpha\beta} +\sigma^{\alpha\beta} \hat{x}\over x^2}
  \left({\lambda^n\over 2}\right)^{ab}
\nonumber \\ & & + {1\over 3^2 2^{10}} \langle\bar{q}q\rangle
\langle g^2_c G^2\rangle
  x^2 \sigma^{\alpha\beta} \left({\lambda^n\over 2}\right)^{ab}
-  {1\over 192}\langle\bar{q}g_c\sigma\cdot Gq\rangle
  \sigma^{\alpha\beta} \left({\lambda^n\over 2}\right)^{ab}
\nonumber \\ & & +  {im_q\over 768}\langle\bar{q}g_c\sigma\cdot
Gq\rangle
  \left( \hat{x} \sigma^{\alpha\beta} +\sigma^{\alpha\beta} \hat{x} \right)
  \left({\lambda^n\over 2}\right)^{ab}
+  {i e_q\over 32\pi^2} F_{\alpha\beta}
  { \hat{x} \sigma^{\alpha\beta} +\sigma^{\alpha\beta} \hat{x} \over x^2 }
  \delta^{ab}
\nonumber \\ & & -  {e_q\over 24} \chi \langle\bar{q}q\rangle
   F_{\alpha\beta} \sigma^{\alpha\beta} \delta^{ab}
+  {ie_q m_q\over 96} \chi \langle\bar{q}q\rangle F_{\alpha\beta}
  \left( \hat{x} \sigma^{\alpha\beta} +\sigma^{\alpha\beta} \hat{x} \right)
   \delta^{ab}
\nonumber \\ & & +  {e_q \over 288} \langle\bar{q}q\rangle
F_{\alpha\beta}
  \left( x^2 \sigma^{\alpha\beta} - 2 x_\rho x^\beta \sigma^{\beta\alpha}
  \right) \delta^{ab}
\nonumber \\ & & +  {e_q \over 576} \langle\bar{q}q\rangle
F_{\alpha\beta}
  \left[ x^2 (\kappa+\xi) \sigma^{\alpha\beta}
  - x_\rho x^\beta (2\kappa-\xi) \sigma^{\beta\alpha} \right] \delta^{ab}
\nonumber \\ & & -  {e_q \over 16} \langle\bar{q}q\rangle
   \left( \kappa F_{\alpha\beta}
   - {i\over 4} \xi \epsilon_{\alpha\beta\mu\nu} F^{\mu\nu} \right)
  \left({\lambda^n\over 2}\right)^{ab}
+ \mbox{higher order terms}. \label{prop}
\end{eqnarray}
We use the convention $\epsilon^{0123}=+1$ in this work.

\end{widetext}

In addition to the standard vacuum condensates,
the vacuum susceptibilities induced by the external field are defined by
\begin{equation}
\begin{array}{r}
\langle\bar{q} \sigma_{\mu\nu} q\rangle_F \equiv
e_q \chi \langle\bar{q}q\rangle F_{\mu\nu}, \\
\langle\bar{q} g_c G_{\mu\nu} q\rangle_F \equiv
e_q \kappa \langle\bar{q}q\rangle F_{\mu\nu}, \\
\langle\bar{q} g_c \epsilon_{\mu\nu\rho\lambda} G^{\rho\lambda}
\gamma_5 q\rangle_F \equiv i e_q \xi \langle\bar{q}q\rangle
F_{\mu\nu}.
\end{array}
\end{equation}
Note that $\chi$ has the dimension of GeV$^{-2}$, while $\kappa$ and
$\xi$ are dimensionless.

\begin{figure*}[p]
\centerline{\epsfig{file=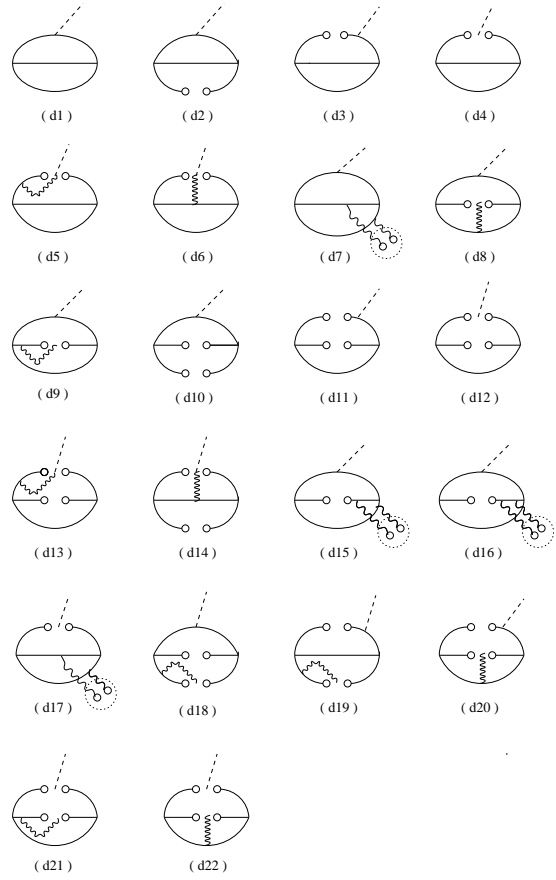,width=7.2cm}} \vspace{1cm}
\vspace*{-1cm}
\caption{Non-mass diagrams considered for the octet baryon magnetic moments.}
\label{xmag}
\end{figure*}

\begin{figure*}[p]
\centerline{\epsfig{file=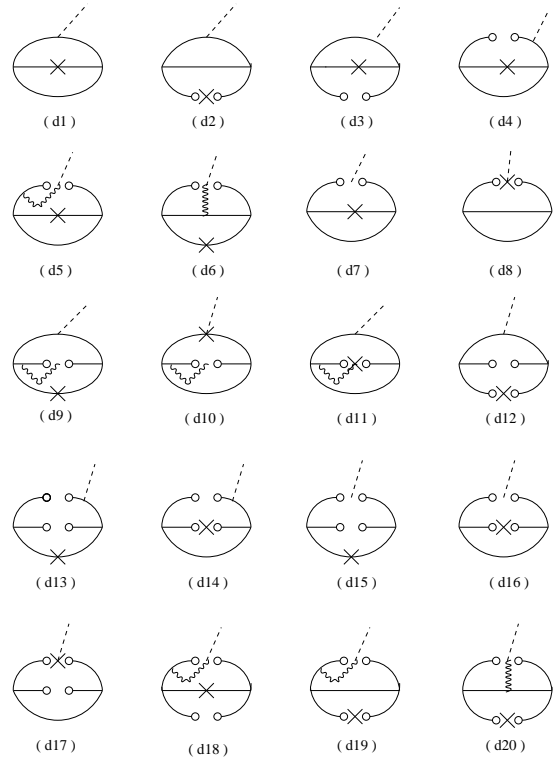,width=7.2cm}} \vspace{1cm}
\vspace*{-1.5cm}
\caption{Diagrams considered for the strange quark mass corrections
to the octet baryon magnetic moments.} \label{xmagm}
\end{figure*}

With the above elements in hand, it is straightforward to evaluate
the correlation function by
substituting the quark propagator into the various master formulae.
We keep terms to first order in the external field and in the strange quark
mass. Terms up to dimension 8 are considered.
The algebra is extremely tedious. Each term in the master formula is a product
of three copies of the quark propagator. There are hundreds of such terms
over various color permutations.
The calculation can be organized by diagrams (similar to Feynmann diagrams)
in Fig.~\ref{xmag} and Fig.~\ref{xmagm}.
Note that each diagram is only generic and all possible color permutations are understood.
The QCD side has the same tensor structure as the phenomenological side and
the results can be organized according to the same 3 independent structures.
%
%
%

\section{QCD Sum Rules}
\label{qcdsr}

Once we have both the QCD side (LHS) and the phenomenological side (RHS),
we can derive the sum rules by matching both sides. Since there are three independent
tensor structures, three sum rules can be constructed.
We denote these tensor structures by the following short-hand notation
\begin{eqnarray}
\mbox{WE}_1 &=&  F^{\mu \nu } (\hat p\sigma _{\mu \nu } + \sigma _{\mu \nu } \hat p),\\
\mbox{WO}_1 &=&  F^{\mu \nu } \sigma _{\mu \nu },\\
\mbox{WO}_2 &=&  F^{\mu \nu } i(p_\mu \gamma _\nu - p_\nu \gamma _\mu )\hat p
\end{eqnarray}
The sum rule from $\mbox{WE}_1$ involves only dimension-even condensates,
so we call this sum rule chiral-even.
The sum rule from both $\mbox{WO}_1$ and $\mbox{WO}_2$ involves only dimension-odd condensates,
so we call them chiral-odd.
Note that in previous works~\cite{Ioffe84,Chiu86} the dimension of the tensor
structures, rather than the dimension of the condensates,
was used to refer to the sum rules. The two names are opposite.

Now we are ready to collect all of the QCD sum rules.
At the structure $\mbox{WE}_1$, all of the sum rules can be expressed in the following form
\begin{widetext}
\begin{equation}
\begin{array}{l}
 {\rm{c}}_{\rm{1}} L^{-4/9} E_2(w) M^4  + {\rm{c}}_{\rm{2}} m_s\chi a  L^{-26/27}E_1(w) M^2+{\rm{c}}_{\rm{3}} \chi a^2 L^{-4/27}E_0(w)  + {\rm{c}}_{\rm{4}} bL^{-4/9} E_0(w) + ({\rm{c}}_{\rm{5}}  + {\rm{c}}_{\rm{6}} )m_s aL^{-4/9}E_0(w)
 \\+ ({\rm{c}}_{\rm{7}}  + {\rm{c}}_{\rm{8}} )a^2 L^{4/9} \frac{1}{{M^2 }}
  + {\rm{c}}_{\rm{9}} \chi m_0 ^2 a^2 L^{ -18/27} \frac{1}{{M^2 }} + {\rm{c}}_{\rm{10}} m_s m_0 ^2 aL^{ - 26/27} \frac{1}{{M^2
  }}\\=- \tilde \lambda _N ^2 [\frac{{\mu_N }}{{M^2 }} + A]e^{ - M_N ^2 /M^2
},
 \end{array}
\label{we1}
\end{equation}
\end{widetext}
where the coefficients differ from member to member.
The quark condensate, gluon condensate, and
the mixed condensate are
\begin{equation}
a=-(2\pi)^2\,\langle\bar{u}u\rangle, \hspace{2mm} b=\langle g^2_c\,
G^2\rangle, \hspace{2mm} \langle\bar{u}g_c\sigma\cdot G
u\rangle=-m_0^2\,\langle\bar{u}u\rangle.
\end{equation}
The quark charge factors $e_q$ are given in units of electric charge
\begin{equation}
e_u=2/3, \hspace{4mm} e_d=-1/3, \hspace{4mm} e_s=-1/3.
\end{equation}
Note that we choose to keep the quark charge factors explicit in the
sum rules. The advantage is that it can facilitate the study of
individual quark contribution to the magnetic moments. The parameters $f$ and $\phi$
account for the flavor-symmetry breaking of the strange quark in the
condensates and susceptibilities:
\begin{equation}
f={ \langle\bar{s}s\rangle \over \langle\bar{u}u\rangle } ={
\langle\bar{s}g_c\sigma\cdot G s\rangle \over
   \langle\bar{u}g_c\sigma\cdot G u\rangle },
\hspace{4mm} \phi={ \chi_s \over \chi }={ \kappa_s \over \kappa }={
\xi_s \over \xi }.
\end{equation}
The anomalous dimension corrections of the interpolating fields and the various
operators are taken into account in the leading logarithmic
approximation via the factor
\begin{equation}
L^\gamma=\left[{\alpha_s(\mu^2) \over \alpha_s(M^2)}\right]^\gamma
=\left[{\ln(M^2/\Lambda_{QCD}^2) \over \ln(\mu^2/\Lambda_{QCD}^2)}
\right]^\gamma,
\end{equation}
where $\mu=500$ MeV is the renormalization scale and $\Lambda_{QCD}$
is the QCD scale parameter. As usual, the pure excited state
contributions are modeled using terms on the OPE side surviving
$M^2\rightarrow \infty$ under the assumption of duality, and are
represented by the factors
\begin{equation}
E_n(w)=1-e^{-w^2/M^2}\sum_n{(w^2/M^2)^n \over n!},
\end{equation}
where $w$ is an effective continuum threshold and it is in principle
different for different sum rules and we will treat it as a free
parameter in the analysis.

Also, $\tilde \lambda _N ^2$ is the rescaled
current coupling ${\tilde \lambda _N ^2 \equiv \lambda^2/4}$.

We only need to carry out four separate calculations for $\Sigma^+$(uus),
$\Sigma^0$(uds), $\Xi^+$(ssd) and $\Lambda$(uds). The QCD sum rules
for other members can be obtained from them by making appropriate
substitutions specified below.

\begin{widetext}

For $\Sigma^+$ at $\mbox{WE}_1$:
\begin{equation}
\begin{array}{*{20}l}
c_1= \frac{1}{16 }[(1 + \beta)^2 c_s-2(3 + 2 \beta + 3 \beta^2) c_u];\\
c_2=\frac{1}{16 }(1 + \beta) [ (-1 + 3 \beta) c_s f_s \phi-2(1 + 3 \beta) c_u];\\
c_3=\frac{1}{12} (-1 + \beta) [(1 + \beta) c_s f_s \phi - 2 c_u (1 -
\beta + f_s + \beta f_s)];
\\
c_4=-\frac{1}{384 }[(3 - 2 \beta + 3 \beta^2) c_s+2(11 + 6 \beta + 11 \beta^2) c_u];\\
c_5=\frac{1}{24}[(3 - 18 \beta + 3 \beta^2 - 2 f_s + 2 \beta^2 f_s) c_s+(-5 + 2 \beta + 27 \beta^2 - 9 f_s + 6 \beta f_s + 27 \beta^2 f_s)c_u];\\
c_6=-\frac{1}{24 }(-1 + \beta)[ (1 + \beta) c_s f_s \phi (8 \kappa - \xi)+ 2c_u(4 \kappa + 6 \beta \kappa + \xi)];\\
 c_7=\frac{1}{72 }(-1 + \beta)[ c_s (-3 + 3 \beta + 2 f_s + 2 \beta f_s) + 2c_u (-1 + \beta + 8 f_s + 8 \beta f_s)];\\
c_8=-\frac{1}{72 }(-1 + \beta) [ (1 + \beta) c_s f_s \phi (8 \kappa
- \xi)
\\\;\;\;\;\;+c_u (-2 \kappa + 2 \beta \kappa + 10 f_s \kappa +
10 \beta f_s \kappa + \xi - \beta \xi +    f_s \xi + \beta f_s \xi)]
;\\c_9=-\frac{1}{288 }[(7(1 + \beta) c_s f_s \phi-c_u (3 - 3 \beta + f_s + \beta f_s)];\\
 c_{10}=-\frac{1}{96
 }[(1 - 14 \beta + \beta^2) c_s+(-23 - 2 \beta + 13 \beta^2) c_u (1 + f_s)].
\end{array}\label{sWE1}
\end{equation}

For $\Xi^0$ at $\mbox{WE}_1$:
\begin{equation}
\begin{array}{*{20}l}
   \begin{array}{l}
c_1= \frac{1}{16 }[(1 + \beta)^2 c_u-2(3 + 2 \beta + 3 \beta^2) c_s];\\
c_2=\frac{1}{16 }(1 + \beta) [ (-1 + 3 \beta)c_u -2(1 + 3 \beta) c_s f_s \phi];\\
c_3=\frac{1}{12} (-1 + \beta) [(1 + \beta) c_u f_s- 2 c_s f_s \phi
(f_s - \beta f_s + 1+ \beta)];
\\
c_4=-\frac{1}{384 }[(3 - 2 \beta + 3 \beta^2) c_u+2(11 + 6 \beta + 11 \beta^2) c_s];\\
c_5=\frac{1}{24}[((3 - 18 \beta + 3 \beta^2)f_s - 2 + 2 \beta^2 ) c_s +((-5 + 2 \beta + 27 \beta^2)f_s - 9 + 6 \beta + 27 \beta^2 )c_s];\\
c_6=-\frac{1}{24 }(-1 + \beta)[ (1 + \beta) c_u (8 \kappa - \xi)+ 2 c_s f_s \phi(4 \kappa + 6 \beta \kappa + \xi)];\\
 c_7=\frac{1}{72 }(-1 + \beta)[ c_u f_s ((-3 + 3 \beta)f_s + 2 + 2 \beta) + 2c_s f_s((-1 + \beta)f_s + 8 + 8 \beta)];\\
c_8=-\frac{1}{72 }(-1 + \beta) [ (1 + \beta) c_u f_s  (8 \kappa -
\xi)
\\\;\;\;\;\;+c_s f_s \phi (-2 \kappa f_s + 2 \beta \kappa f_s+ 10  \kappa +
10 \beta \kappa + \xi - \beta f_s\xi +    f_s \xi + \beta \xi)]
;\\c_9=-\frac{1}{288 }[(7(1 + \beta) c_u f_s-c_s f_s \phi ((3 - 3 \beta)f_s + 1+ \beta)];\\
 c_{10}=-\frac{1}{96
 }[(1 - 14 \beta + \beta^2) c_u +(-23 - 2 \beta + 13 \beta^2) c_s (1 + f_s)].
\end{array}  \\ \\
\end{array}\label{xWE1}
\end{equation}

For $\Lambda$ at $\mbox{WE}_1$:
\begin{equation}
\begin{array}{*{20}l}
   \begin{array}{l}
c_1=-\frac{1}{48} ((-1 + \beta)^2 (c_d +c_u)+ (13 + 10 \beta + 13
\beta^2) c_s )
;\\c_2=\frac{1}{16} (1 + \beta) ((-1 + \beta) (c_d +c_u) - (1 + 5 \beta) c_s f_s \phi)
;\\c_3=\frac{1}{72} (-1 + \beta) ((c_d +c_u) (1 - \beta + f_s + 5 \beta f_s) - 2 (5 + \beta) c_s f_s \phi)
;\\c_4=-\frac{1}{1152}((17 + 2 \beta + 17 \beta^2) (c_d +c_u) + (41 + 26
\beta + 41 \beta^2) c_s)
;\\c_5=\frac{1}{144} (c_s (-78 + \beta (84 - 16 f_s) + \beta^2 (210 -
4 f_s) + 20 f_s) \\ \;\;\;\;\;+ (-1 + \beta) (c_d +c_u)(7 + 3 f_s +
\beta (41 + 33 f_s)) )
;\\c_6=-\frac{1}{72} (-1 + \beta) (c_s f_s \phi (8 (1 + 2 \beta)
\kappa + (5 + \beta) \xi) +
   (c_d +c_u) ((20 + 22 \beta) \kappa - (1 + 2 \beta) \xi))
;\\c_7=\frac{1}{216} (-1 + \beta) (c_s (39 + 33 \beta - 10 f_s - 2
\beta f_s) +
   (c_d +c_u) (1 - \beta + 4 f_s + 20 \beta f_s) )
;\\c_8=-\frac{1}{432} (-1 + \beta) (2 c_s f_s \phi (8 (1 + 2 \beta)
\kappa + (5 + \beta) \xi) \\ \;\;\;\;\;+
  (c_d +c_u) (2 (19 + 17 \beta + f_s + 5 \beta f_s) \kappa - (1 - \beta + f_s +
         5 \beta f_s) \xi) )
;\\c_9=-\frac{1}{576} (-1 + \beta) ((c_d +c_u) (5 + 3 \beta + 3 f_s + 7 \beta
f_s) - 2 (5 + \beta) c_s f_s \phi)
;\\c_{10}=\frac{1}{192} ((62 - 4 \beta - 34 \beta^2) c_s - (-7 - 10
\beta + 5 \beta^2) (c_d +c_u) (1 +
      f_s)).
\end{array}
\end{array}\label{lWE1}
\end{equation}

At the structure $\mbox{WO}_1$, the sum rules can be expressed in the following form
\begin{equation}
\begin{array}{l}
{\rm{c}}_{\rm{1}} m_s L^{-8/9}E_2(w) M^4  + {\rm{c}}_{\rm{2}} \chi
aL^{ - 16/27} E_2(w) M^4  + ({\rm{c}}_{\rm{3}} {\rm{ + c}}_{\rm{4}}
)a E_1(w) M^2  + {\rm{c}}_{\rm{5}} m_s \chi a^2 L^{-16/27}E_0(w) \\+
{\rm{c}}_{\rm{6}} \chi ab L^{-16/27}E_0(w)+
({\rm{c}}_{\rm{7}}+{\rm{c}}_{\rm{8}}) m_s a^2 \frac{1}{{M^2 }}+
{\rm{c}}_{\rm{9}} ab \frac{1}{{M^2 }}
\\=-\tilde \lambda_N ^2 M_N \left[ \frac{2\mu_N
}{M^2 } + \frac{\mu_N -1}{M_N^2 } +A \right]e^{ - M_N ^2 /M^2}.
 \end{array}
\label{wo1}
\end{equation}

For $\Sigma^+$ at $\mbox{WO}_1$:
\begin{equation}
\begin{array}{*{20}l}
   \begin{array}{l}
  c_1=-\frac{1}{2 } (-1 + \beta) ((1 + \beta) c_s - 2 c_u)
 ; \\c_2=-\frac{1}{24 } (-1 + \beta) [(-1 + \beta)c_s f_s \phi+ 18 (1 + \beta)
  c_u]
;\\c_3=\frac{1}{24}(-1 + \beta) [c_s (-6 - 6 \beta - f_s + \beta f_s) -6c_u (2 + 2 \beta - f_s + \beta
f_s)]
;\\c_4=\frac{1}{96 }(-1 + \beta)[(-1 + \beta) c_s f_s \phi (14 \kappa - 13 \xi)-18(1 + \beta) c_u (2 \kappa + \xi)]
;\\c_5=-\frac{1}{12}[ (3 + 2 \beta + 3 \beta^2) c_s f_s \phi-(3 + 4 \beta + 9 \beta^2) c_u (1 + f_s)]
;\\c_6=\frac{1}{576 }(-1 + \beta)[ (-1 + \beta) c_s f_s \phi -  18(1 + \beta)
c_u]
;\\c_7=-\frac{1}{36 }[ c_s (-3 + 3 \beta^2 + 5 f_s - 2 \beta f_s + 5 \beta^2 f_s)- c_u (3 + 4
\beta + 9 \beta^2 - 3 f_s + 10 \beta f_s + 9 \beta^2 f_s)]
;\\c_8=-\frac{1}{144}[ c_s f_s \phi (4 \kappa + 8 \beta \kappa + 4 \beta^2 \kappa + \xi -
   10 \beta \xi + \beta^2 \xi)
   \\ \;\;\;\;\;- c_u (-(3 + 4 \beta + 9 \beta^2) (1 + f_s) \xi+4 (2 \beta + 6 \beta^2 + 3 f_s + 2 \beta f_s + 3 \beta^2 f_s) \kappa)];\\
   c_9=-\frac{1}{576}[3 c_s-c_u (-3 + 3 \beta^2 - f_s + 2 \beta f_s)].
 \end{array}
\end{array}\label{sWO1}
\end{equation}
For $\Xi^0$ at $\mbox{WO}_1$:
\begin{equation}
\begin{array}{*{20}l}
   \begin{array}{l}
  c_1=-\frac{1}{2 } (-1 + \beta) ((1 + \beta) c_u - 2 c_s);
  \\c_2=-\frac{1}{24 } (-1 + \beta) [(-1 + \beta)c_u+ 18 (1 + \beta)
  c_s f_s \phi];
\\c_3=\frac{1}{24}(-1 + \beta) [c_u ((-6 - 6 \beta)f_s - 1 + \beta ) -6c_s ((2 + 2 \beta)f_s - 1+ \beta)];
\\c_4=\frac{1}{96 }(-1 + \beta)[(-1 + \beta) c_u (14 \kappa - 13 \xi)-18(1 + \beta) c_s f_s \phi (2 \kappa + \xi)];
\\c_5=-\frac{1}{12}[ (3 + 2 \beta + 3 \beta^2) c_u-(3 + 4 \beta + 9 \beta^2) c_s f_s \phi (1 + f_s)];
\\c_6=\frac{1}{576 }(-1 + \beta)[ (-1 + \beta) c_u -  18(1 + \beta)
c_s f_s \phi];
\\c_7=-\frac{1}{36 }[ c_u f_s ((-3 + 3 \beta^2 )f_s+ 5 - 2 \beta  + 5 \beta^2 )- c_s f_s ((3 + 4
\beta + 9 \beta^2)f_s - 3 + 10 \beta  + 9 \beta^2)];
\\c_8=-\frac{1}{144}[ c_u f_s (4 \kappa + 8 \beta \kappa + 4 \beta^2 \kappa + \xi -
   10 \beta \xi + \beta^2 \xi)
   \\ \;\;\;\;\;- c_s f_s \phi (-(3 + 4 \beta + 9 \beta^2) (1 + f_s) \xi+4 ((2 \beta + 6 \beta^2 )f_s+ 3  + 2 \beta + 3 \beta^2) \kappa)];\\
   c_9=-\frac{1}{576}[3 c_u-c_s ((-3 + 3 \beta^2)f_s - 1 + 2 \beta )].
 \end{array}
\end{array}\label{xWO1}
\end{equation}

For $\Lambda$ at $\mbox{WO}_1$:
\begin{equation}
\begin{array}{*{20}l}
   \begin{array}{l}
c_1=-\frac{1}{6} (-1 + \beta) ((c_d +c_u) (1+ 2 \beta ) - (5 +
\beta) c_s );
\\c_2=-\frac{1}{72} (-1 +
   \beta) ((7 + 11 \beta) (c_d +c_u)  + (37 + 35 \beta) c_s f_s \phi);
\\c_3=-\frac{1}{72} (-1 + \beta) ((c_d +c_u) (14 + \beta (22 - 3 f_s) + 3
f_s)  +
   c_s (-30 - 6 \beta + 35 f_s + 37 \beta f_s));
\\c_4=\frac{1}{288} (-1 + \beta) ((c_d +c_u)(2 (-23 + 5 \beta) \kappa + (17
- 35 \beta) \xi)  -
   c_s f_s \phi (58 \kappa + 86 \beta \kappa + 49 \xi + 23 \beta \xi));
\\c_5=\frac{1}{24} ((-1 + \beta^2) (c_d +c_u)(1 + f_s) +
   2 (5 + 6 \beta + 13 \beta^2) c_s f_s \phi);
\\c_6=-\frac{1}{1728}((-1 +
    \beta) ((11 + 7 \beta) (c_d +c_u)  + (35 +
       37 \beta) c_s f_s \phi));
\\c_7=\frac{1}{216} (-(-1 + \beta) (c_d +c_u) (-7 + \beta - f_s + 13 \beta
f_s) +
   c_s (-30 + 34 f_s + 4 \beta (6 + 7 f_s) + \beta^2 (6 + 82 f_s)));
\\c_8=\frac{1}{864} (2 c_s f_s \phi (4 (7 + 10 \beta + 19 \beta^2)
\kappa - (11 + 26 \beta +35 \beta^2) \xi) \\ \;\;\;\;\;+ (-1 +
      \beta)(c_d +c_u) (4 (-4 - 2 \beta + 5 f_s + 7 \beta f_s) \kappa - (-5 +
         11 \beta) (1 + f_s) \xi) );
\\c_9=\frac{1}{3456}((-1 + \beta) (2 (5 + \beta) c_s + (c_d +c_u) (5 + \beta
+ 11 f_s + 13 \beta f_s) )).
\end{array}
\end{array}\label{lWO1}
\end{equation}

At the structure $\mbox{WO}_2$, the sum rules can be expressed in the following form
\begin{equation}
\begin{array}{l}
 {\rm{c}}_{\rm{1}} m_s L^{-8/9} E_1(w) M^2  + {\rm{c}}_{\rm{2}} \chi aL^{ - 16/27} E_1(w) M^2  + ({\rm{c}}_{\rm{3}} {\rm{ + c}}_{\rm{4}} )a E_0(w) + {\rm{c}}_{\rm{5}} m_s \chi a^2 L^{-16/27}E_0(w) + {\rm{c}}_{\rm{6}} m_0 ^2 aL^{-4/9} \frac{1}{{M^2 }} \\
  + c_{\rm{7}} \chi abL^{-16/27} \frac{1}{{M^2 }} + ({\rm{c}}_{\rm{8}}  + {\rm{c}}_{\rm{9}} )m_s a^2  \frac{1}{{M^4 }}{\rm{ + c}}_{{\rm{10}}} ab\frac{1}{{M^4 }} \\
 =  -\frac{\tilde \lambda _N ^2}{M_N} [\frac{{2(\mu_N-1) }}{{M^2 }} + A]e^{ - M_N ^2 /M^2
}.
 \end{array}
\label{wo2}
\end{equation}

For $\Sigma^+$ at $\mbox{WO}_2$:
\begin{equation}
\begin{array}{*{20}l}
   \begin{array}{l}
c_1=-\frac{1}{2 }(-1 + \beta) [(1 + \beta) c_s-2c_u];
\\c_2=-\frac{1}{6 }(-1 + \beta)^2 c_s f_s \phi;\\
c_3=-\frac{1}{4 }(-1 + \beta) [ c_s (2 + 2 \beta - f_s + \beta
f_s)+2c_u (-2 - 2 \beta - f_s + \beta f_s)];
\\c_4=\frac{1}{16 } (-1 + \beta) [c_s f_s \phi+2(1 + \beta) c_u](2 \kappa - \xi);
\\c_5=-\frac{1}{3 } (-1 + \beta) [2c_s f_s \phi+ c_u (-1 - \beta - f_s + \beta f_s)];
\\c_6=\frac{1}{24 }(-1 + \beta) [7(1 + \beta) c_s+ c_u (-1 - \beta - 3 f_s + 3 \beta f_s)];
\\c_7=-\frac{1}{144 }(-1 + \beta)^2 c_s f_s \phi;
\\c_8=\frac{1}{18} [c_s (-3 + 3 \beta^2 + 5 f_s - 2 \beta f_s + 5 \beta^2 f_s) - c_u (3 + 4
\beta + 9 \beta^2 - 3 f_s + 10 \beta f_s + 9 \beta^2 f_s)];
\\c_9=\frac{1}{72} [c_s f_s \phi ((1 - 10 \beta + \beta^2) \xi+ 4 (1 + \beta)^2 \kappa)
\\\;\;\;\;\;- c_u
(4 (2 \beta + 6 \beta^2 + 3 f_s + 2 \beta f_s + 3 \beta^2 f_s)
\kappa-(3 + 4 \beta +
9 \beta^2) (1 + f_s) \xi)];\\
c_{10}=-\frac{1}{288} (-1 + \beta) ((1 + \beta) c_s-c_u (-3 - 3
\beta - f_s + \beta f_s)].
 \end{array}
\end{array}\label{sWO2}
\end{equation}
For $\Xi^0$ at $\mbox{WO}_2$:
\begin{equation}
\begin{array}{*{20}l}
   \begin{array}{l}
c_1=-\frac{1}{2 }(-1 + \beta) [(1 + \beta) c_u-2c_s];
\\c_2=-\frac{1}{6 }(-1 + \beta)^2 c_u;\\
c_3=-\frac{1}{4 }(-1 + \beta) [ c_u f_s((2 + 2 \beta)f_s - 1 +
\beta)+2c_s f_s(-(2 + 2 \beta)f_s - 1 + \beta )];
\\c_4=\frac{1}{16 } (-1 + \beta) [c_s f_s \phi+2(1 + \beta) c_u](2 \kappa - \xi);
\\c_5=-\frac{1}{3 } (-1 + \beta) [2c_u f_s+ c_s f_s \phi(-(1 + \beta)f_s -1 + \beta)];
\\c_6=\frac{1}{24 }(-1 + \beta) [7(1 + \beta) c_u f_s+ c_s  (-(1 + \beta)f_s - 3  + 3 \beta )];
\\c_7=-\frac{1}{144 }(-1 + \beta)^2 c_u;
\\c_8=\frac{1}{18} [c_u f_s ((-3 + 3 \beta^2)f_s + 5  - 2 \beta+ 5 \beta^2 ) - c_s f_s ((3 + 4
\beta + 9 \beta^2)f_s - 3+ 10 \beta + 9 \beta^2 )];
\\c_9=\frac{1}{72} [c_u f_s ((1 - 10 \beta + \beta^2) \xi+ 4 (1 + \beta)^2 \kappa)
\\\;\;\;\;\;- c_s f_s
(4 ((2 \beta + 6 \beta^2)f_s + 3  + 2 \beta  + 3 \beta^2) \kappa-(3
+ 4 \beta +
9 \beta^2) (1 + f_s) \xi)];\\
c_{10}=-\frac{1}{288} (-1 + \beta) ((1 + \beta) c_u -c_s (-(3 + 3
\beta)f_s - 1 + \beta )].
 \end{array}
\end{array}\label{xWO2}
\end{equation}
For $\Lambda$ at $\mbox{WO}_2$:
\begin{equation}
\begin{array}{*{20}l}
   \begin{array}{l}
c_1=-\frac{1}{12} (-1 + \beta) ((c_d+c_u)(1 + 2 \beta) - (5 + \beta)
c_s );
\\c_2=-\frac{1}{36} (-1 + \beta)^2 (2 (c_d+c_u) - c_s f_s \phi);
\\c_3=\frac{1}{24} (-1 + \beta) ((c_d+c_u) (2 + \beta (-6 + f_s) - f_s) +
   + c_s (10 + 3 f_s + \beta (2 + 5 f_s)));
\\c_4=\frac{1}{96} (-1 +
   \beta) ((-1 + 3 \beta) (c_d+c_u) + (5 +
      3 \beta) c_s f_s \phi) (2 \kappa - \xi);
\\c_5=\frac{1}{12} (-1 + \beta) ((c_d+c_u)(-1 + \beta (-1 + f_s) - f_s)  + 4 c_s f_s \phi);
\\c_6=\frac{1}{96} (-1 + \beta) (-2 (5 + \beta) c_s + (c_d+c_u) (3 + 7 \beta
+ 5 f_s + 3 \beta f_s) );
\\c_7=-\frac{1}{864} (-1 + \beta)^2 (2 (c_d+c_u) - c_s f_s \phi);
\\c_8=\frac{1}{216} ((-1 + \beta) (c_d+c_u) (-7 + \beta - f_s + 13 \beta
f_s)  -
   2 c_s (-15 + 17 f_s + 2 \beta (6 + 7 f_s) + \beta^2 (3 + 41 f_s)));
\\c_9=\frac{1}{864} (2 c_s f_s \phi (-4 (7 + 10 \beta + 19 \beta^2)
\kappa + (11 + 26 \beta +
         35 \beta^2) \xi) \\\;\;\;\;\;- (-1 +
      \beta) (c_d+c_u) (4 (-4 - 2 \beta + 5 f_s + 7 \beta f_s) \kappa - (-5 +
         11 \beta) (1 + f_s) \xi) );
\\c_{10}=-\frac{1}{3456}((-1 + \beta) (2 (5 + \beta) c_s + (c_d+c_u) (5 +
\beta + 11 f_s + 13 \beta f_s) ).
 \end{array}
\end{array}\label{lWO2}
\end{equation}

\end{widetext}
It is not necessary to list all the coefficients for all 8 members since
the coefficients for other members of the octet family can be obtained
by appropriate replacements of quark contents in the following way:
\begin{itemize}
\item for proton $p$, replace s quark by d quark in $\Sigma^+$.
\item for neutron $n$, exchange d quark with u quark in proton $p$,
\item for $\Sigma^-$, replace u quark by d quark in $\Sigma^+$,
\item for $\Xi^-$, replace u quark by d quark in $\Xi^0$,
\item for $\Sigma^0$, change corresponding factors in $\Xi^0$.
\end{itemize}
Here the conversions between u and d quarks are achieved by simply
switching their charge factors $e_u$ and $e_d$. Moreover, The
conversions from s quark to u or d quarks involve setting $m_s=0$,
$f=\phi=1$, in addition to the switching of charge factors. This
way, they can also provide additional checks for our calculations.

At this point, we can make some comparisons with previous
calculations in Ref.~\cite{Pasupathy86,Chiu86,Wilson87}.
First, we use general interpolating fields where we can vary $\beta$
to achieve the best match in sum rules.
The previous calculations correspond to a fixed value of $\beta$=-1.
This effect was studied in detail in Ref.~\cite{Derek96}, and it was found that
$\beta=-1.2$ is the optimal value. Most of our results are at $\beta=-1.2$.
Second, we have checked that our sum rules agree with those in the previous calculations
for the most part. For example, for the proton at $\mbox{WE}_1$, we completely agree except
for the $\kappa-2\xi$ term in Eq.~(2.16) in Ref.~\cite{Chiu86}.
In all of our sum rules, we have the combination $2\kappa-\xi$ instead of $\kappa-2\xi$.
For the strange members ($\Sigma$, $\Xi$, $\Lambda$),
they have 8 terms in the OPE, while we have 10 terms.
For the sum rules at structure $\mbox{WO}_1$, they have only 3 terms, while we have 9 terms.
For the sum rules at structure $\mbox{WO}_2$, they have only 4 terms, while we have 10 terms.
Third, they only analyzed the sum rules at structure $\mbox{WE}_1$,
while we will examine all the structures.
Fourth, we use a completely different analysis method.

Before going into the analysis, we would like to point out some
relations among the correlation functions (or OPE) based on
symmetries, which lead to the same relations in the magnetic
moments. In exact SU(3)-flavor symmetry, it is known that the
magnetic moments of the octet family are related by (see, for
example, Ref.~\cite{Coleman})
\begin{equation}
\begin{array}{l}
\mu_{\Sigma^+}=\mu_p, \\
2\mu_{\Lambda}=\mu_n, \\
\mu_{\Sigma^-}+\mu_n=-\mu_p, \\
\mu_{\Xi^-}=\mu_{\Sigma^-}, \\
\mu_{\Xi^0}=\mu_n.
\end{array}
\label{sym}
 \end{equation}
These relations are borne out in the OPE of our sum rules if SU(3)-flavor symmetry
is enforced. They are only approximately true since SU(3)-flavor symmetry is broken by the
strange quark. Here we have the advantage of studying the symmetry-breaking
effects since the terms are explicit in our QCD sum rules.
%
%
\section{Sum Rule Analysis}
\label{ana}
The sum rules for magnetic moments have the generic form of
OPE - ESC = Pole + Transition, or
\begin{equation}
\Pi_{mag}(QCD,\beta,w,M^2) = \tilde{\lambda}_N^2 \left({\mu_N \over M^2} + A\right) e^{-M_N^2/M^2},
\end{equation}
where $QCD$ represents all the QCD input parameters. The task then
becomes: given the function $\Pi_{mag}$ with known QCD input
parameters and the ability to vary $\beta$, find the
phenomenological parameters (magnetic moment $\mu_{N}$, transition
strength $A$, coupling strength $\tilde{\lambda}_N^2$, and continuum
threshold $w$) by matching the two sides over some region in the
Borel mass $M$. A $\chi^2$ minimization is best suited for this purpose.
It turns out that there are too many fit parameters for this
procedure to be successful in general. To alleviate the situation,
we employ the corresponding mass sum rules which have a similar
generic form of OPE - ESC = Pole, or
\begin{equation}
\Pi_{mass}(QCD,\beta,w_1,M^2) = \tilde{\lambda}_N^2 e^{-M_N^2/M^2},
\end{equation}
which shares some of the common parameters. Note that the continuum threshold may not be
the same in the two sum rules.
By taking the ratio of the two equations, we are left with
\begin{equation}
{\Pi_{mag}(QCD,\beta,w,M^2) \over \Pi_{mass}(QCD,\beta,w_1,M^2)} =
\frac{\mu_N}{M^2}+A. \label{ratio}
\end{equation}
This is the form we are going to implement.
By plotting the two sides as a function of $1/M^2$, the slope
will be the magnetic moment and the intercept the transition strength.
The linearity (or deviation from it) of the
left-hand side gives an indication of OPE convergence and
the role of excited states.
The two sides are expected to match for a good sum rule. This way of
matching the sum rules has two advantages. First, the slope, which is the magnetic moment
of interest, is usually better determined than the intercept.
Second, by allowing the possibility of different continuum thresholds,
we ensure that both sum rules stay in their valid regimes.

We use the chiral-even mass sum rules in Ref.~\cite{Lee02} which are
listed here in the same notation,
\begin{equation}
\begin{array}{l}
p_1  L^{-4/9} E_3 (w_1)M^6
                 + p_2 b  L^{-4/9} E_1 (w _1 )M^2
                + p_3 m_s a L^{4/9}
                \\+ p_4 a^2 L^{4/9}
                 + p_5 a^2 k_v L^{4/9}
                + p_6 m_0^2 a^2 L^{-2/27}\frac{1}{M^2}\\=\tilde \lambda _N ^2 e^{ - M_N ^2 /M^2
                }.
\end{array}
 \end{equation}
The coefficients for N are:
\begin{equation}
\begin{array}{*{20}l}
   \begin{array}{l}
   p_1=\frac{1}{64} (5 +\beta  + 5\beta ^2);
   \\p_2= \frac{1}{256} (5 +\beta  + 5\beta ^2);
   \\p_3=0; \\
   p_4= \frac{1}{24}(7 -2\beta  -5\beta );\\
   p_5=0;\\
   p_6=  - \frac{1}{96}(13 - 2\beta  - 11\beta ^2 ).
\end{array}
\end{array}
\label{massp}
 \end{equation}
For $\Lambda$:
\begin{equation}
\begin{array}{*{20}l}
   \begin{array}{l}
p_1=\frac{1}{64} (5+2 \beta+5 \beta^2);
 \\ p_2= \frac{1}{256} (5+2 \beta+5 \beta^2);
 \\p_3=\frac{1}{96}((20-15f_s)-(16+6f_s)\beta-(4+15 f_s)\beta^2);
 \\ p_4=\frac{1}{96}((4f_s-5-6t)+(4+4f_s)\beta+(4 f_s+1+6 t)\beta^2);
 \\ p_5= \frac{1}{72} ((10 f_s+11)+(2-8 f_s) \beta-(2 f_s+13) \beta^2);
 \\ p_6= \frac{1}{288} ((-16 f_s-23)+(8 f_s-2) \beta+(8 f_s+25) \beta^2).
 \end{array}
\end{array}
\label{masslam}
 \end{equation}
For $\Sigma$:
\begin{equation}
\begin{array}{*{20}l}
   \begin{array}{l}
p_1=\frac{1}{64} (5+2 \beta+5 \beta^2);
 \\ p_2=\frac{1}{256}(5+2 \beta+5 \beta^2);
 \\ p_3=\frac{1}{32}((12-5 f_s)-2 f_s \beta-(12+5f_s) \beta^2);
\\ p_4=- \frac{1}{94}((4 f_s+21+18 t)+4 f_s \beta +(4 f_s-21-18t) \beta^2);
 \\ p_5=\frac{1}{24} ((6 f_s+1)-2 \beta-(6 f_s-1) \beta^2);
 \\ p_6=-\frac{1}{96}((12 f_s+1)-2 \beta-(12 f_s-1)\beta^2).
 \end{array}
\end{array}
\label{masssig}
 \end{equation}
For $\Xi$:
\begin{equation}
\begin{array}{*{20}l}
   \begin{array}{l}
p_1=  \frac{1}{64} (5+2 \beta+5 \beta^2);
 \\ p_2= \frac{1}{256} (5+2 \beta+5 \beta^2);
 \\ p_3= \frac{3}{16} ((2-f_s)-2 f_s \beta-(2+f_s) \beta^2);
\\ p_4=-\frac{1}{96}((15-f_s+18 t)-10 f_s \beta-(15+f_s+18 t) \beta^2);
 \\ p_5= \frac{1}{24} f_s ((f_s+6)-2 f_s \beta+(f_s-6) \beta^2);
 \\ p_6= - \frac{1}{94} f_s ((f_s+12)-2 f_s \beta+(f_s-12) \beta^2).
\end{array}
\end{array}\label{massxi}
 \end{equation}
The function t is defined as $t\equiv \ln{\frac{M^2}{\mu^2}}-\gamma_{EM}$
with $\gamma_{EM}\approx 0.577$ the Euler-Mascheroni constant.
Use this mass sum rule for the coupling constant for magnetic
moment sum rules, and move to the left side, then only the magnetic
moment $\mu_N$ and the transition strength $A$ dependence  are on
the right hand side.

We use the Monte-Carlo procedure first introduced in Ref.~\cite{Derek96} to
carry out the search which allows a rigorous error analysis.
In this method, the entire phase-space of the input QCD parameters
is explored simultaneously, and is mapped into uncertainties in the phenomenological
parameters.
This lead to more realistic uncertainty estimates than traditional approaches.

First, a set of randomly-selected, Gaussianly-distributed condensates
are generated with a assigned uncertainties. Here we give 10\% for
the uncertainties of input parameters, and this number can be
adjusted to test the sensitivity of the QCD parameters.  Then the OPE
is constructed in the Borel window with evenly distributed points $M_j$.
Note that the uncertainties in the OPE are not uniform
throughout the Borel window. They are larger at the lower end where
uncertainties in the higher-dimensional condensates dominate. Thus,
it is crucial that the appropriate weight is used in the calculation
of $\chi^2$. For the OPE obtained from the k'th set of QCD
parameters, the $\chi^2$ per degree of freedom is
\begin{equation}
{\chi^2_k\over N_{DF}}= \sum^{n_{\scriptscriptstyle
B}}_{j=1} { [\Pi^{\scriptscriptstyle OPE}_k(M^2_j, \beta, w, w_1)
-\Pi^{\scriptscriptstyle Phen}_k(M^2_j,\mu,A)]^2 \over
(n_B-n_p)\;\sigma^2_{\scriptscriptstyle OPE}(M_j) },
\end{equation}
where $\Pi^{\scriptscriptstyle OPE}$ refers to the LHS of Eq.~(\ref{ratio}) and
$\Pi^{\scriptscriptstyle Phen}$ its RHS.
The integer $n_p$ is the number of phenomenological search parameters.
In this work, $n_B$=51 points were used along the
Borel axis. The procedure is repeated for many QCD parameter sets,
resulting in distributions for phenomenological fit parameters, from
which errors are derived. In practice, 200 configurations are
sufficient for getting stable uncertainties. We used about 2000 sets
to resolve more subtle correlations among the QCD parameters
and the phenomenological fit parameters.
This means that each sum rule is fitted 2000 times to arrive at the final results.

The QCD input parameters are given as follows.
The condensates are taken as $a=0.52\; GeV^3$,$ b=1.2\; GeV^4$,$ m^2_0=0.72\; GeV^2$.
For the factorization violation parameter, we use $\kappa_v=2.0$.
The QCD scale parameter is restricted to $\Lambda_{QCD}=0.15$ GeV.
The vacuum susceptibilities have
been estimated in studies of nucleon magnetic
moments~\cite{Ioffe84,Chiu86,Lee98b,MAM}, but the values vary
depending on the method used. We use $\chi=-6.0\; GeV^{-2}$
and $\kappa=0.75$, $\xi=-1.5.$
Note that $\chi$ is almost an order of magnitude larger than
$\kappa$ and $\xi$, and is the most important of the three. The
strange quark parameters are placed at
$ m_s=0.15\; GeV$, $f=0.83$, $\phi=0.60$~\cite{Pasupathy86,Lee98b}.
These input parameters are just central values.
We will explore sensitivity to these parameters by assigning
uncertainties to them in the Monte-Carlo analysis.
%

\section{Result and Discussion}
\label{res}

\begin{table*}[thb] 
\caption{Results for the magnetic moment of octet baryons from the
QCD sum rule in Eq.~(\protect\ref{we1}) (structure $\mbox{WE}_1$).
The seven columns correspond to, from left to right: particle, $\beta$
value, Borel region in which the two sides of the QCD sum rule are
matched, continuum threshold, transition strength,
extracted magnetic moment in unit of nuclear magnetons, and experimental value.
The errors are derived from 2000 samples in the Monte-Carlo analysis with 10\% uncertainty
on all QCD input parameters.}
\label{tabwe1}
 \begin{tabular*}{0.75\textwidth}{@{\extracolsep{\fill}}lcccccc}
\hline\hline
                & $\beta$  & Region           & $w$       & A              &$\mu_{\scriptscriptstyle B}$ &Exp. \\
                      &          & (GeV)         & (GeV)         & (GeV$^{-2}$)     & $(\mu_{\scriptscriptstyle N})$&$(\mu_{\scriptscriptstyle N})$\\ \hline
$p$ \hspace{1mm} &-1.2   & 0.7 to 0.9 & 1.40&1.46  $\pm$ 0.34      & 3.01  $\pm$ 0.24   &   2.79  \\
    \hspace{1mm} &-0.4   & 0.8 to 1.2 & 1.60&0.74  $\pm$ 0.13      & 2.82  $\pm$ 0.26   &   2.79  \\
$n$
\hspace{1mm}      & -1.2  &  0.7 to 1.1 & 1.40&-0.19  $\pm$ 0.09     &-1.97  $\pm$ 0.15   &  -1.91  \\
$\Lambda$
\hspace{1mm} & -1.2  &  1.1 to  1.2 & 1.60  &  -0.45  $\pm$ 0.05   & -0.56  $\pm$ 0.15   &  -0.61\\
$\Sigma^+$
\hspace{1mm}     &   -1.2   & 1.1 to 1.3 & 1.85  &  0.56  $\pm$ 0.06     &2.31  $\pm$ 0.25      &2.45\\
 $\Sigma^0$
     \hspace{1mm} & -1.2    &1.0 to 1.6& 1.80&0.07  $\pm$ 0.02        &0.69  $\pm$ 0.07  &    0.65 \\
 $\Sigma^-$
 \hspace{1mm} &  -1.2  &  1.3 to 1.8 & 1.80&-0.08  $\pm$ 0.01    &  -1.16  $\pm$ 0.10   &  -1.16\\
$\Xi^0$
\hspace{1mm} &   -1.2 &   1.6 to 1.9 & 2.15   & -0.16  $\pm$ 0.01    &  -1.15  $\pm$ 0.05    & -1.25 \\
 $\Xi^-$
\hspace{1mm} &  -1.2 &   1.1 to 1.4 & 2.00&-0.34  $\pm$ 0.02   &    -0.64  $\pm$ 0.06    & -0.65\\
\hline\hline\\
\end{tabular*}
\end{table*}
%

We have 24 sum rules in total to analyze: 3 for each member of the octet.
For each sum rule, we have in principle 5 parameters to determine:
$\mu$, $A$, $w$, $w_1$, $\beta$.
But a search treating all five parameters as free does not work because there is
not enough information in the OPE.
In fact, the freedom to vary $\beta$ can be used as an advantage to yield the optimal
match.  We find that $\beta=-1.2$ gives the best match in most cases. This agrees with
the value suggested in Ref.~\cite{Derek96}.
One exception is the proton: we found a better solution at $\beta=-0.4$ than at $\beta=-1.2$.
Another parameter that can be used to our advantage is the continuum threshold $w_1$
for the corresponding mass sum rule. We fix it to the value
that gives the best solution to the mass sum rule independently.
The following values for $w_1$ are used:
for the nucleon, $w_1=1.44$ GeV; for $\Lambda$, $w_1=1.60$ GeV; for $\Sigma$, $w_1=1.66$ GeV;
for $\Xi$, $w_1=1.82$ GeV.
In this way the magnetic moment sum rule and the mass sum rule can stay in their respective
valid Borel regimes.
This leaves us with three parameters: $\mu$, $A$, $w$.
Unfortunately, a three-parameter search is either unstable or returns values for $w$
smaller than the particle mass, an unphysical situation. Again we think this is a symptom
of insufficient information in the OPE.
So we are forced to fix the continuum threshold $w$ that corresponds to the best match
for the central values of the QCD parameters.

%
\begin{figure}[h]
    \begin{center}
      \epsfig{file=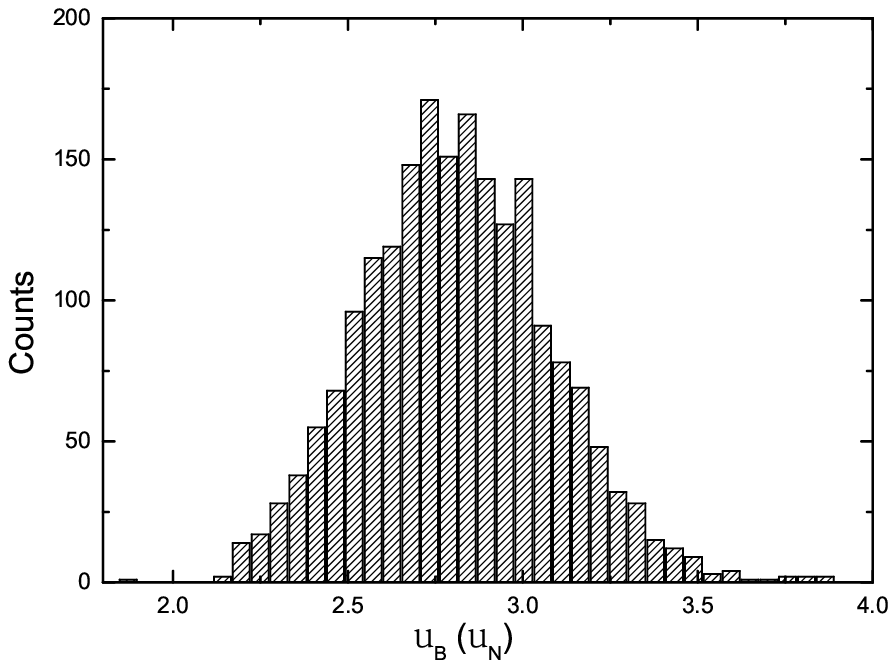, scale=0.80}
      \end{center}
\vspace{-1.5cm}
    \begin{center}
      \epsfig{file=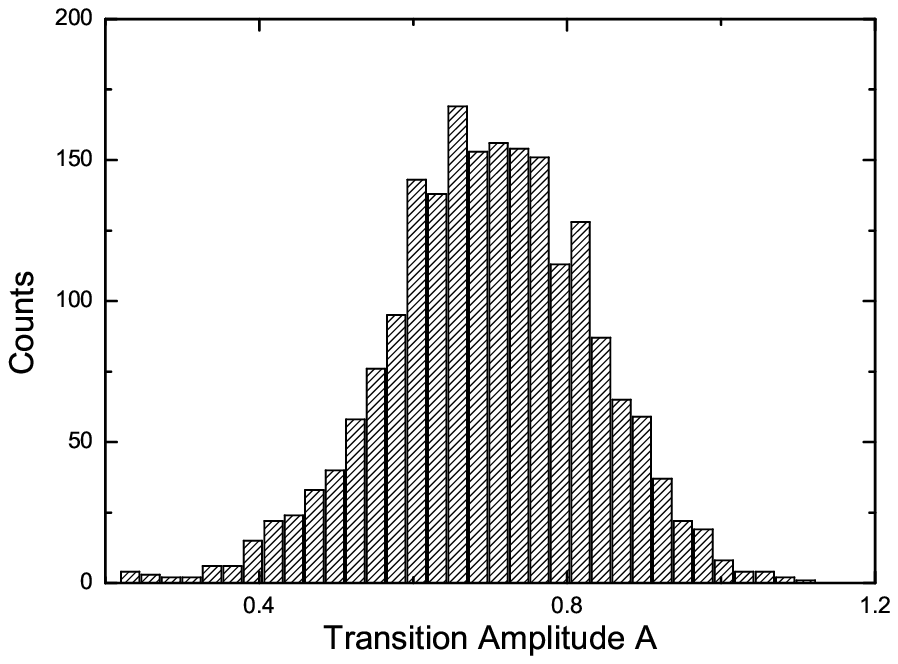, scale=0.80}
      \end{center}
\vspace{-1.0cm}
  \caption{Histogram for the proton magnetic moment (top) and
transition amplitude (bottom) obtained from Monte-Carlo fits of
Eq.~\protect\ref{we1}) at $\mbox{WE}_1$ for 2000 QCD parameter sets.
They are based on 10\% uncertainty given to all the QCD input
parameters.}
  \label{histo}
\end{figure}
\subsection{The Sum Rule at $\mbox{WE}_1$}
The results determined this way at the $\mbox{WE}_1$ structure are
displayed in Table~\ref{tabwe1}.
The Borel window is determined by the following two criteria:
OPE convergence which gives the lower bound, and ground-state dominance
which gives the upper bound. It is done iteratively. For each value of $\beta$, we adjust
the Borel window until the best solution is found.
We see that our calculated magnetic moments agree with experiment fairly well
within error bars.

We stress that the errors are derived from Monte-Carlo distributions
which give the most realistic estimation of the uncertainties. An
example of such distributions is given in Fig.~\ref{histo}. We see
that they are roughly Gaussian distributions. The central value is
taken as the average, and the error is one standard deviation of the
distribution. We found about 10\% accuracy for the magnetic moments
in our Monte-Carlo analysis, resulting from 10\% uniform uncertainty
in all the QCD input parameters. Of course, the uncertainties in the
QCD parameters can be non-uniform. For example, we tried the
uncertainty assignments (which are quite conservative) in
Ref.~\cite{Derek96}, and found about 30\% uncertainties in our
output.

\begin{figure}[tbh]
    \begin{center}
      \epsfig{file=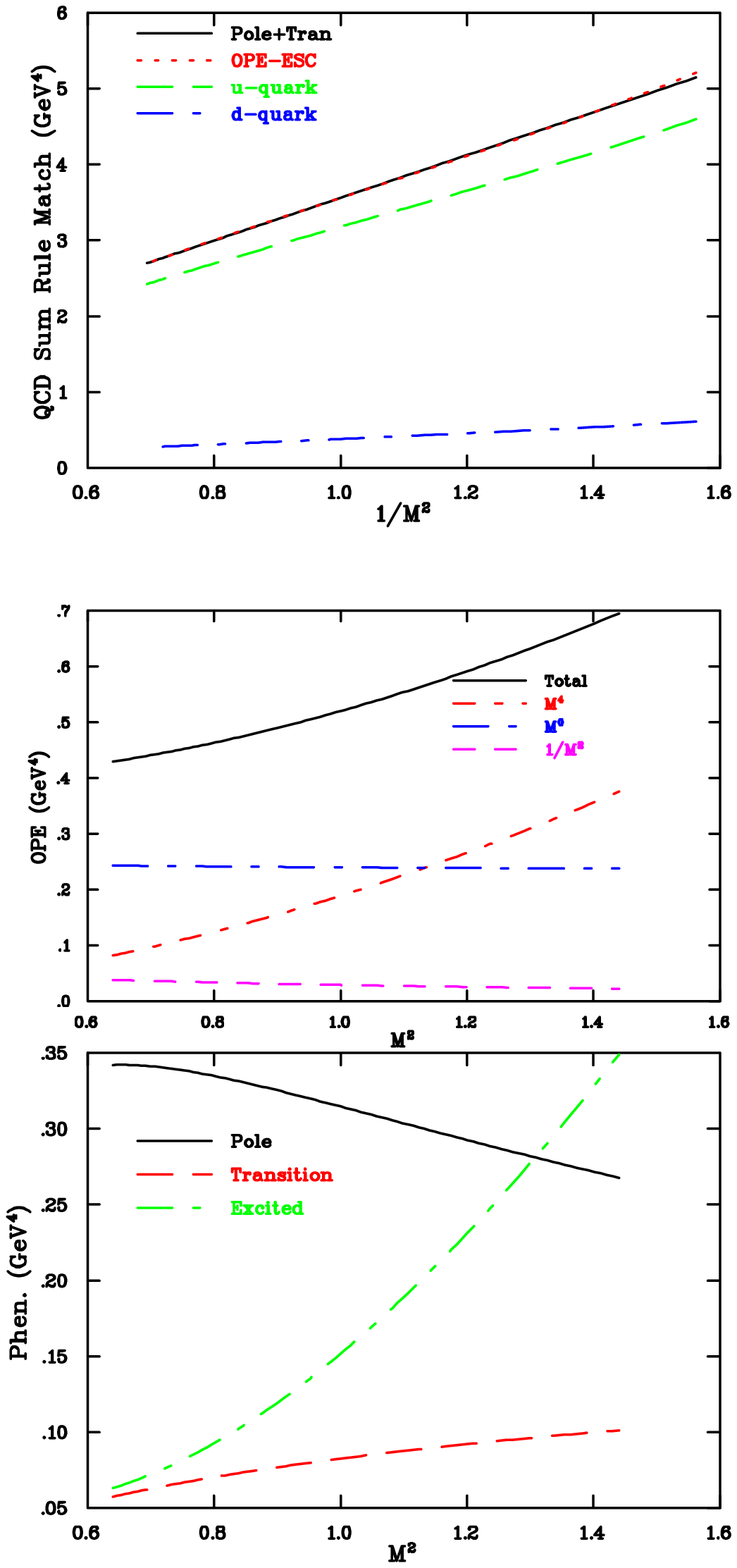, scale=0.65}
      \caption{Analysis of the QCD sum rule in Eq.~(\protect\ref{we1})
(structure $\mbox{WE}_1$) for the proton at $\beta=-0.4$ according
to Eq.~(\protect\ref{ratio}). In the top, the pole plus transition
terms (solid lines) are compared against the OPE minus the
excited-state contributions (dashed lines) as a function of $1/M^2$
(the two should match for an ideal sum rule). Also plotted are the
individual contributions from u (long-dashed lines) and d
(dot-dashed lines) quarks. In the middle, the total in the OPE side
and its various terms are plotted as a function of $M^2$. In the
bottom, the 3 terms in the phenomenological side: pole (solid),
transition (long-dashed), and excited (dot-dashed) are plotted as a
function of $M^2$.}
      \label{pwe1}
    \end{center}
\end{figure}
%

To gain a better appreciation on how the QCD sum rules produce the
results, we show Fig.~\ref{pwe1}, using the proton as an example.
There are three graphs in this figure to give three different
aspects of the analysis. The first graph shows how the two sides of
Eq.~(\protect\ref{ratio}) match over the Borel window, which should
be linear as a function of $1/M^2$ according to the right-hand side
of this equation. Indeed, we observe excellent linear behavior from
the OPE side (LHS). The match is almost perfect (barely
distinguishable between the solid and dotted lines). The slope gives
the magnetic moment $\mu$, and the intercept give the transition
contribution $A$. We find that the inclusion of $A$ is important in
producing the best match. Also plotted are the individual
contributions from u and d quarks. We see that for the proton, the
u-quark contribution is the dominant one, which is expected because
it is doubled-represented in the proton ($uud$). We define the slope
from an individual quark contribution as the effective magnetic
moment of that quark in the particle.

The second graph in Fig.~\ref{pwe1} shows how the various terms in
the OPE contribute to determination of magnetic moments. The $M^0$
term, which contains the contributions from the condensates $\chi
a^2$ and $b$, plays an important role. It is the leading
contribution in the region below $M^2<1.2\; \mbox{GeV}^2$. For this
reason, the sum rule at $\mbox{WE}_1$ is expected to have good
spectral properties. Indeed this is confirmed in the third graph
where we plot the three terms in the phenomenological side (pole,
transition, and excited) as a function of $M^2$. The ground-state
pole is dominant (over 70\% of the RHS at the low end of the Borel
window). The excited-state contribution starts small, then grows
with $M^2$, as expected from the continuum model. The transition
contribution is small in this sum rule. It is consistently smaller
than the excited-state contribution and has a weak dependence on the
Borel mass.

\subsection{The Sum Rule at $\mbox{WO}_1$}
%
%
\begin{table*}[thb] 
\caption{Similar to Table~\protect\ref{tabwe1}, but for the
QCD sum rule in Eq.~(\protect\ref{wo1}) (structure $\mbox{WO}_1$).}
\label{tabwo1}
\begin{tabular*}{ 0.75\textwidth}{@{\extracolsep{\fill}}lcccccc}
\hline\hline
                & $\beta$  & Region           & $w$       & A                &$\mu_{\scriptscriptstyle B}$ &Exp. \\
                       &          & (GeV)         & (GeV)         & (GeV$^{-1}$)     & $(\mu_{\scriptscriptstyle N})$&$(\mu_{\scriptscriptstyle N})$\\ \hline
$p$ \hspace{1mm} &   -0.8   &  1.4 to 1.6   & 1.50             &   1.03$\pm$0.16      & 2.67 $\pm$0.16                &2.79  \\
 $n$
\hspace{1mm}      &   -1.2    &1.2 to 1.4    &  1.40       &  -0.54 $\pm$0.11       &  -1.70 $\pm$0.11             &-1.91   \\
 $\Lambda$
\hspace{1mm}     & -1.2  & 1.0 to 1.3     & 1.60   & -2.3 $\pm$0.31       &-0.62 $\pm$0.17  &-0.61\\
$\Sigma^+$
     \hspace{1mm} &  -0.6 & 1.5 to 1.7      & 1.60 &1.60 $\pm$0.50      & 2.46 $\pm$0.40  & 2.45 \\
 $\Sigma^0$
 \hspace{1mm} & -1.0  & 1.4 to 1.6      & 1.70 & 0.23 $\pm$0.11     &  0.58 $\pm$0.09  & 0.65\\
 $\Sigma^-$
 \hspace{1mm} & -0.6  & 1.2 to 1.5      & 1.60 &-2.04 $\pm$0.34    & -0.57 $\pm$0.26  &-1.16\\
$\Xi^0$
\hspace{1mm} &  -1.2 & 1.2 to 1.4      &2.10  & -1.52 $\pm$0.21    &-1.27 $\pm$0.15  &-1.25 \\
 $\Xi^-$
\hspace{1mm} &-0.2  & 1.7 to 1.8     &1.90&-0.4 $\pm$0.41      & -0.49 $\pm$0.29   & -0.65\\
\hline\hline\\
\end{tabular*}
\end{table*}
%
Next, we analyze the sum rule in Eq.~(\ref{wo1}) at the structure
$\mbox{WO}_1$, using the same procedure. Table~\ref{tabwo1} displays
the results extracted from this sum rule. The magnetic moments have
larger errors than those from $\mbox{WE}_1$: about 15\% as opposed
to 10\%. The agreement with experiment is reasonable (with the
exception of $\Sigma^-$), but not as good as those from
$\mbox{WE}_1$. We had to search a wider region in $\beta$ to find
the best match. The transition contribution ($A$) are larger for the
strange particles, as well as their errors.

\begin{figure}[tbh]
    \begin{center}
      \epsfig{file=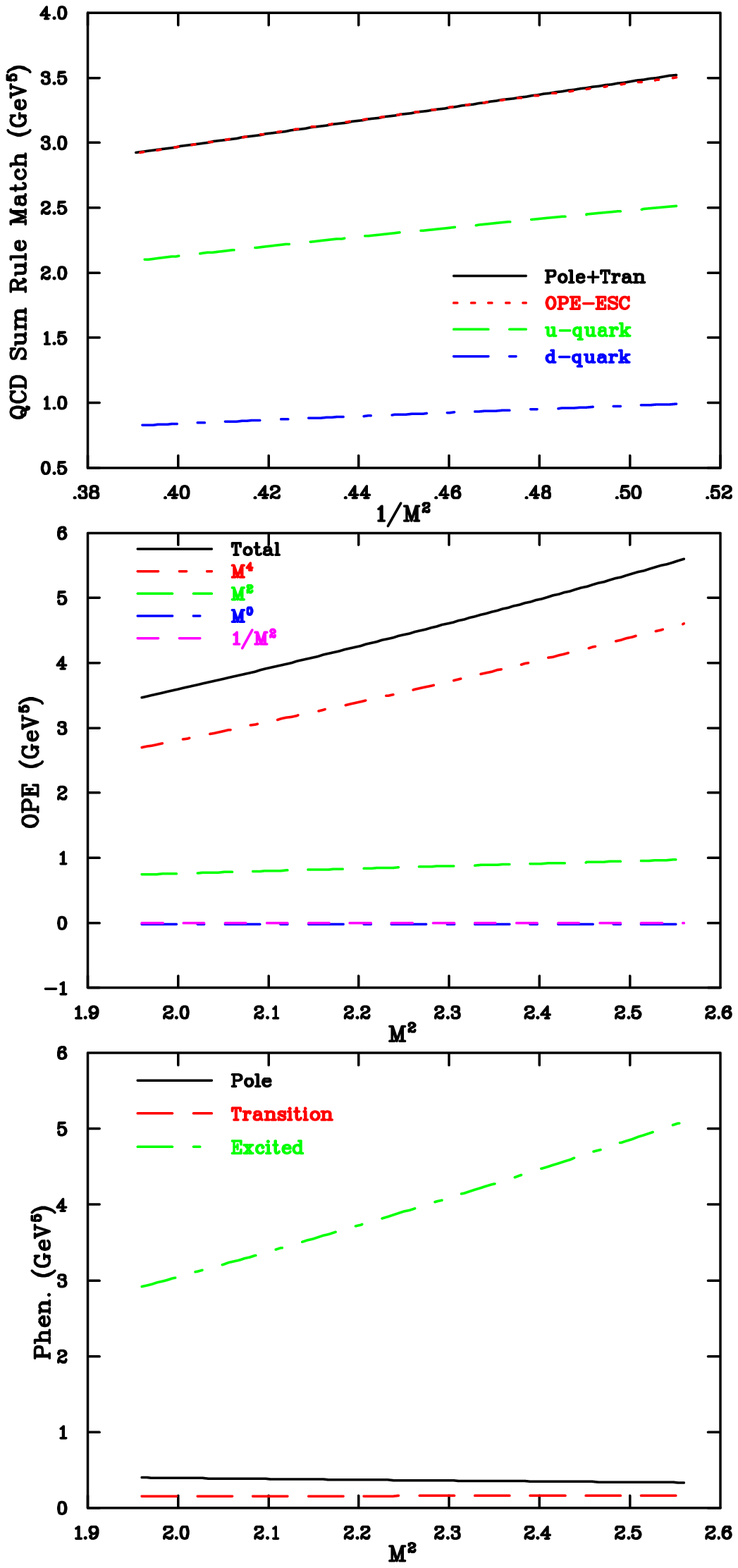, scale=0.65}
  \caption{Similar to Fig.~\protect\ref{pwe1}, but at structure $\mbox{WO}_1$ and $\beta=-1.2$.}
      \label{pwo1}
    \end{center}
\end{figure}
Fig.~\ref{pwo1} shows the details of the analysis in the case of the
proton. The matching is very good, as indicated in the top graph.
The middle graphs shows that the leading contribution in the OPE
($M^4$ term) is $\chi a$, followed by the quark condensate $a$
($M^2$ term). The condensate $\chi a b$ ($M^0$ term) and $ab$
($M^{-2}$ term) are very small in this sum rule. The bottom graph
reveals a surprising result: the excited-state dominates over the
pole and the transition. As a result, this sum rule is less
reliable. This is the reason why the results from this sum rule are
not as good as those from $\mbox{WE}_1$. This sum rule also shows
the importance of checking the individual terms in the
phenomenological side, in addition to looking at the best match of
the two sides. In this case, there is no pole dominance, even though
the leading terms is non-perturbative and the match is almost
perfect.

\subsection{The Sum Rule at $\mbox{WO}_2$}
\begin{figure}[tbh]
    \begin{center}
      \epsfig{file=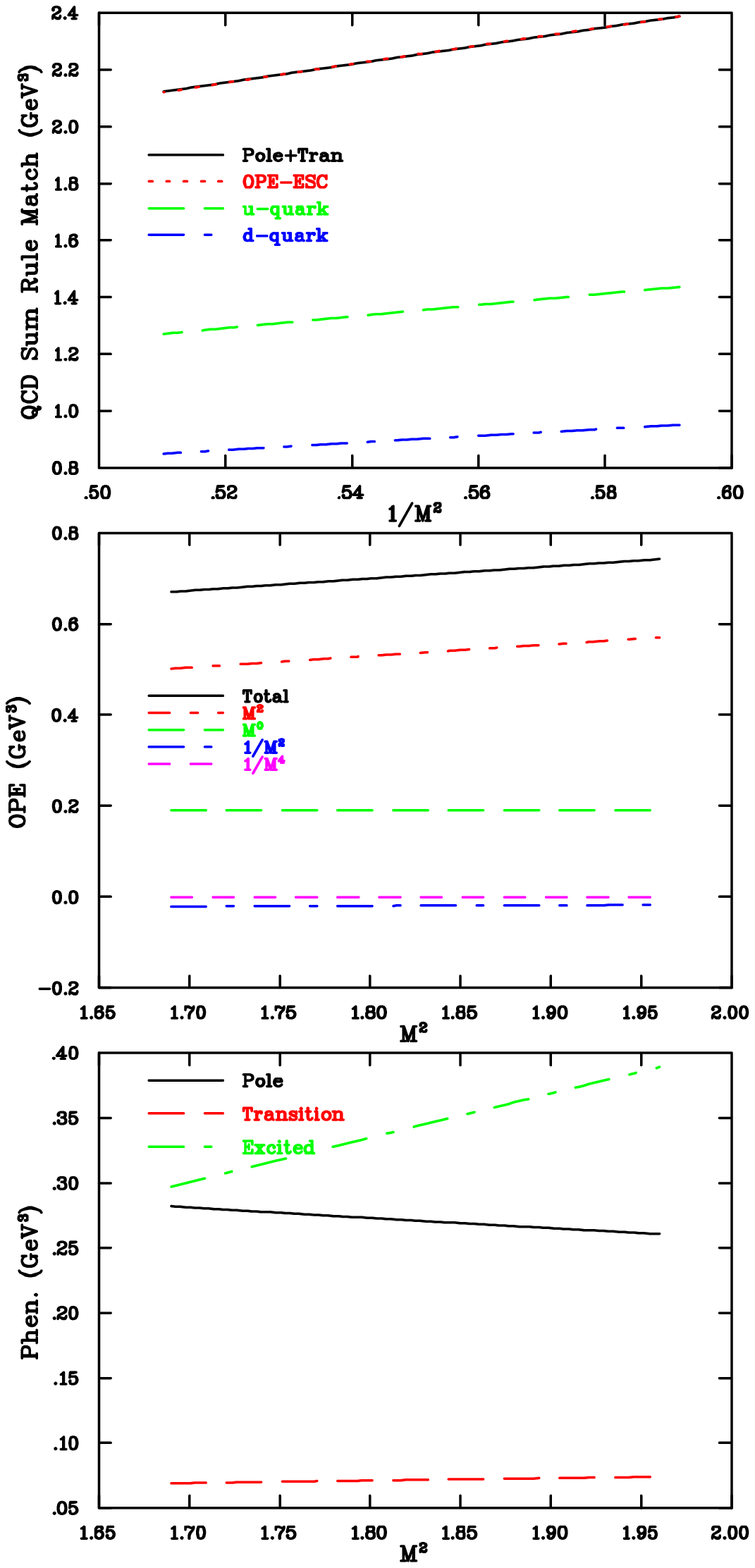, scale=0.65}
  \caption{Similar to Fig.~\protect\ref{pwe1}, but at structure $\mbox{WO}_2$ and $\beta=-1.2$.}
      \label{pwo2}
    \end{center}
\end{figure}

Finally, we present the results from the sum rule in Eq.~(\ref{wo2})
at the structure $\mbox{WO}_2$ in Table~\ref{tabwo1}. The agreement
with experiment is not as good as the other two sum rules. For
example, $\Sigma^-$ and $\Xi^-$ have the wrong sign. Fig.~\ref{pwo2}
shows the details of the analysis for the proton. The matching is
very good, as indicated in the top graph. The middle graph shows
that the leading contribution in the OPE is $M^2$ with a coefficient
of $\chi a$, followed by the $M^0$ term. The $1/M^2$ term is
slightly negative, while the $1/M^4$ term is very small. The bottom
graph shows that the excited-state dominates over the pole and the
transition, like the sum rule from $\mbox{WO}_1$, but the relative
size of the pole is much larger. Since the $\mbox{WO}_2$ sum rule
has power corrections up to $1/M^4$, it is expected to be more
reliable than the $\mbox{WO}_1$ sum rule. But our analysis shows
that this advantage is offset by the smallness of the $1/M^2$ and
$1/M^4$ terms. As a result, the reliability of the $\mbox{WO}_2$ sum
rule is about the same as the $\mbox{WO}_1$ sum rule.
%
%
\begin{table*}[thb] 
\caption{Similar to Table~\protect\ref{tabwe1}, but for the
QCD sum rule in Eq.~(\protect\ref{wo2}) (structure $\mbox{WO}_2$).}
\label{tabwo2}
\begin{tabular*}{ 0.75\textwidth}{@{\extracolsep{\fill}}lcccccc}
\hline\hline
                & $\beta$  & Region           & $w$       & A       &$\mu_{\scriptscriptstyle B}$ &Exp. \\
                       &          & (GeV)         & (GeV)         & (GeV$^{-2}$)   &$(\mu_{\scriptscriptstyle N})$&$(\mu_{\scriptscriptstyle N})$\\ \hline
$p$ \hspace{1mm} &    -1.2  & 1.3  to 1.4   &    1.60      & 0.42 $\pm$0.07                & 2.53 $\pm$0.13    &2.79  \\
 $n$
\hspace{1mm}      & -1.2      & 1.7 to 1.9    &   1.60      &   0.26 $\pm$0.03             &-1.73 $\pm$0.28   &-1.91   \\
$\Lambda$
\hspace{1mm} &-1.2   &  1.3 to 1.4     &1.90  &-0.85 $\pm$0.22     &-0.53 $\pm$0.17  &-0.61\\
 $\Sigma^+$
\hspace{1mm}     & -1.0  & 1.2 to 1.4      & 1.70  &0.21 $\pm$0.05     &  1.50 $\pm$0.11  &2.45\\
 $\Sigma^0$
     \hspace{1mm} &-1.2   & 1.0 to 1.6      & 1.60  &-0.02 $\pm$0.02     & 1.04 $\pm$0.07  & 0.65 \\
 $\Sigma^-$
 \hspace{1mm} & -1.2  & 1.4 to 1.6      & 1.65  &  0.0 $\pm$0.01       &0.61 $\pm$0.06 & -1.16\\
$\Xi^0$
\hspace{1mm} &  -1.2 & 1.6 to 2.0      & 2.1  & -0.08 $\pm$0.03     &-1.20 $\pm$0.23   &-1.25 \\
 $\Xi^-$
\hspace{1mm} &-1.2   & 1.4 to 1.5      &2.2   & 0.22 $\pm$0.09      &0.80 $\pm$0.21  & -0.65\\
\hline\hline
\end{tabular*}
\end{table*}
%

\begin{figure*}[tbh]
  \begin{minipage}[t]{.3\textwidth}
    \begin{center}
      \epsfig{file=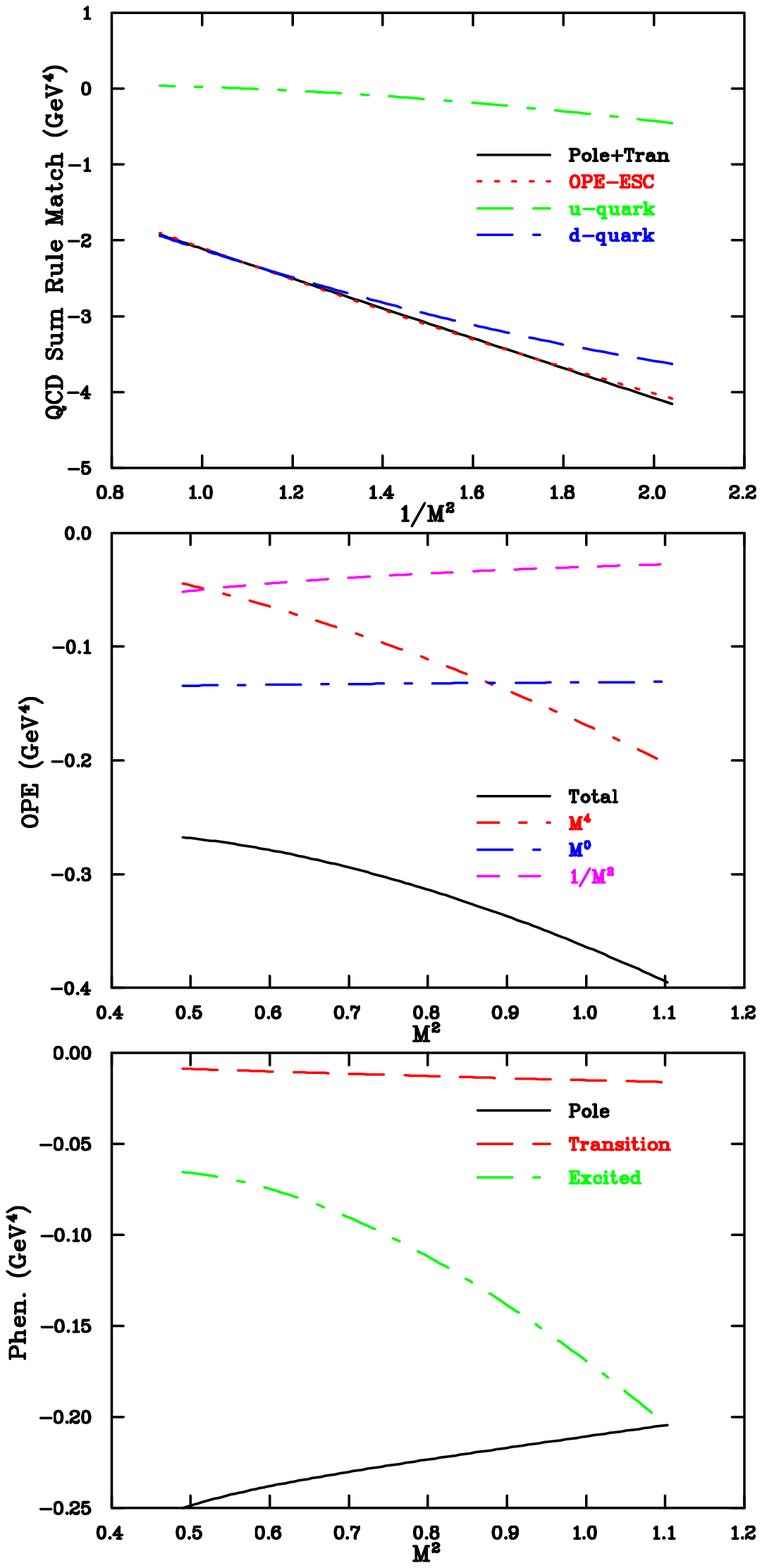, scale=0.45}
\end{center}
  \end{minipage}
  \begin{minipage}[t]{.3\textwidth}
    \begin{center}
      \epsfig{file=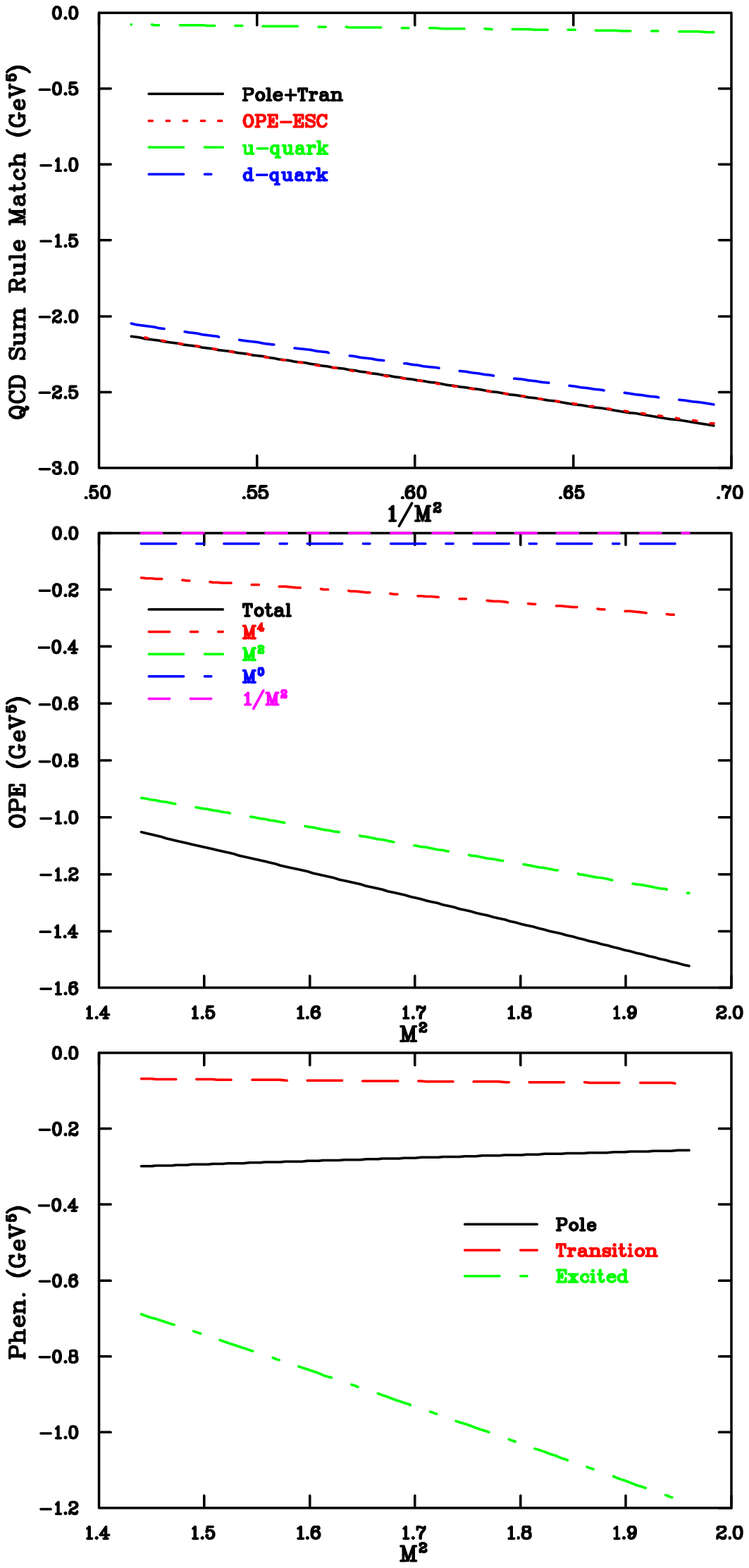, scale=0.45}
      \end{center}
  \end{minipage}
  \begin{minipage}[t]{.3\textwidth}
    \begin{center}
      \epsfig{file=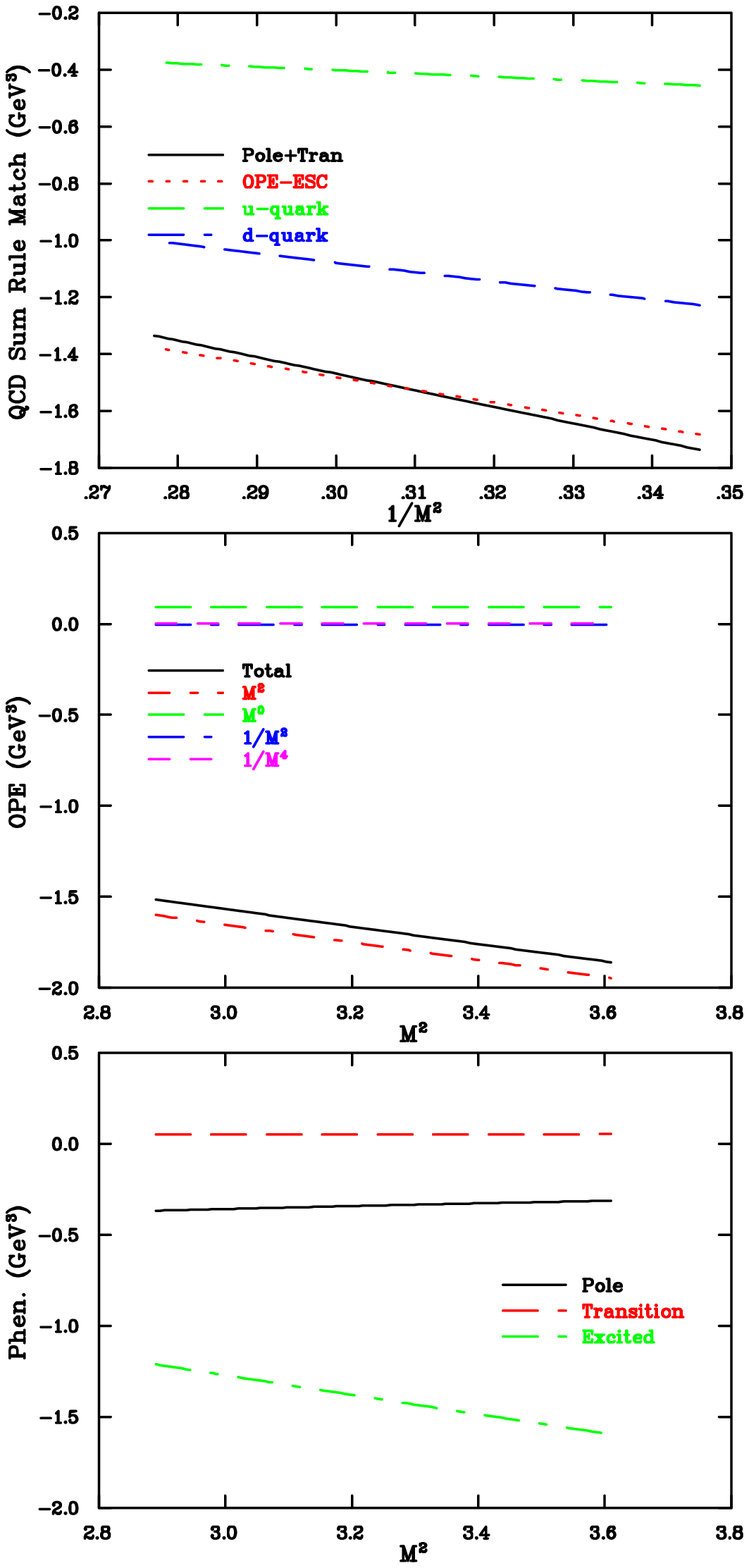, scale=0.45}
      \end{center}
  \end{minipage}
  \caption{Similar to Fig.~\protect\ref{pwe1}, BUT for the neutron at
all three structures: $\mbox{WE}_1$ (left panel), $\mbox{WE}_1$ (middle panel),
$\mbox{WE}_1$ (right panel).}
\label{nfig}
\end{figure*}

\begin{figure*}[tbh]
  \begin{minipage}[t]{.3\textwidth}
    \begin{center}
      \epsfig{file=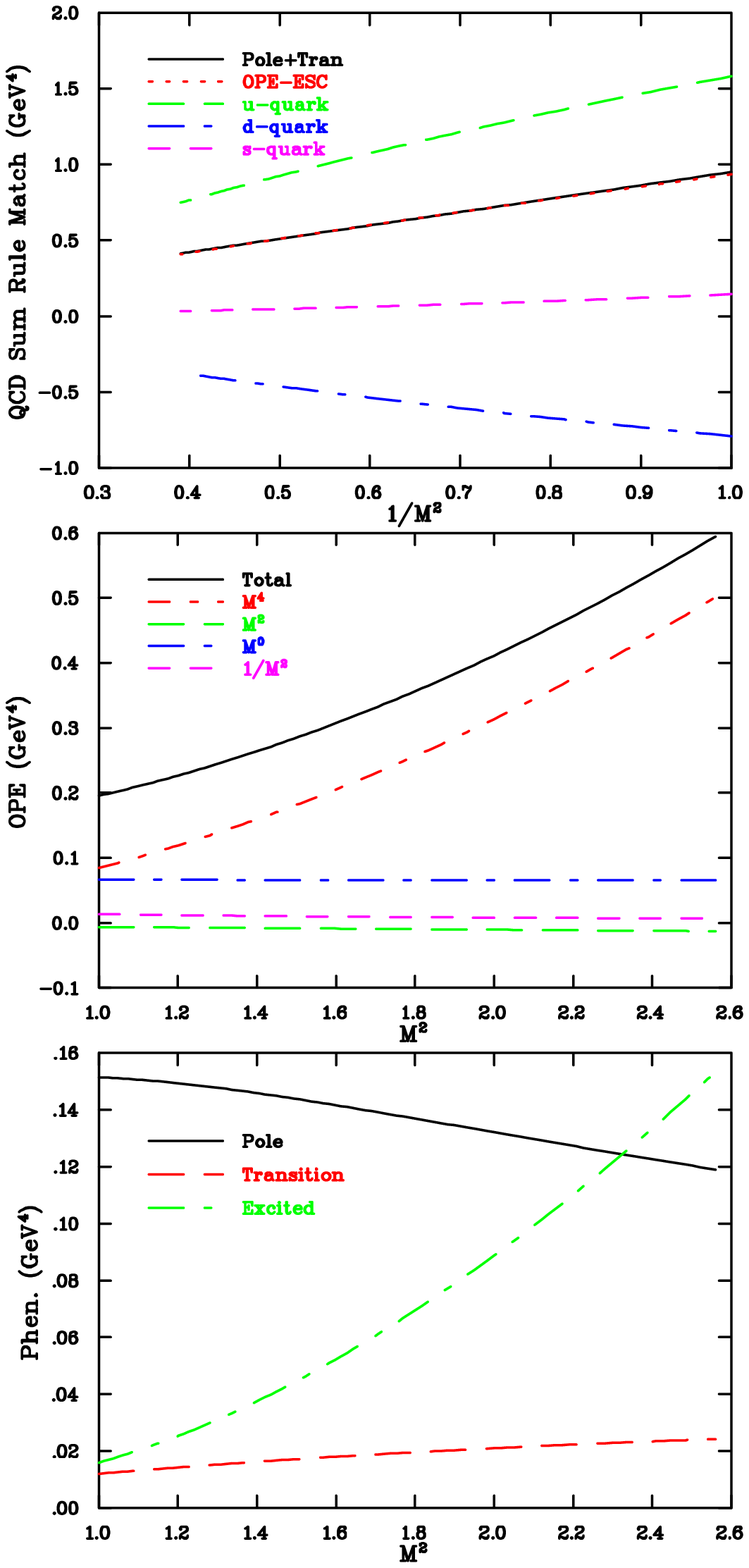, scale=0.45}
      \end{center}
  \end{minipage}
  \begin{minipage}[t]{.3\textwidth}
    \begin{center}
      \epsfig{file=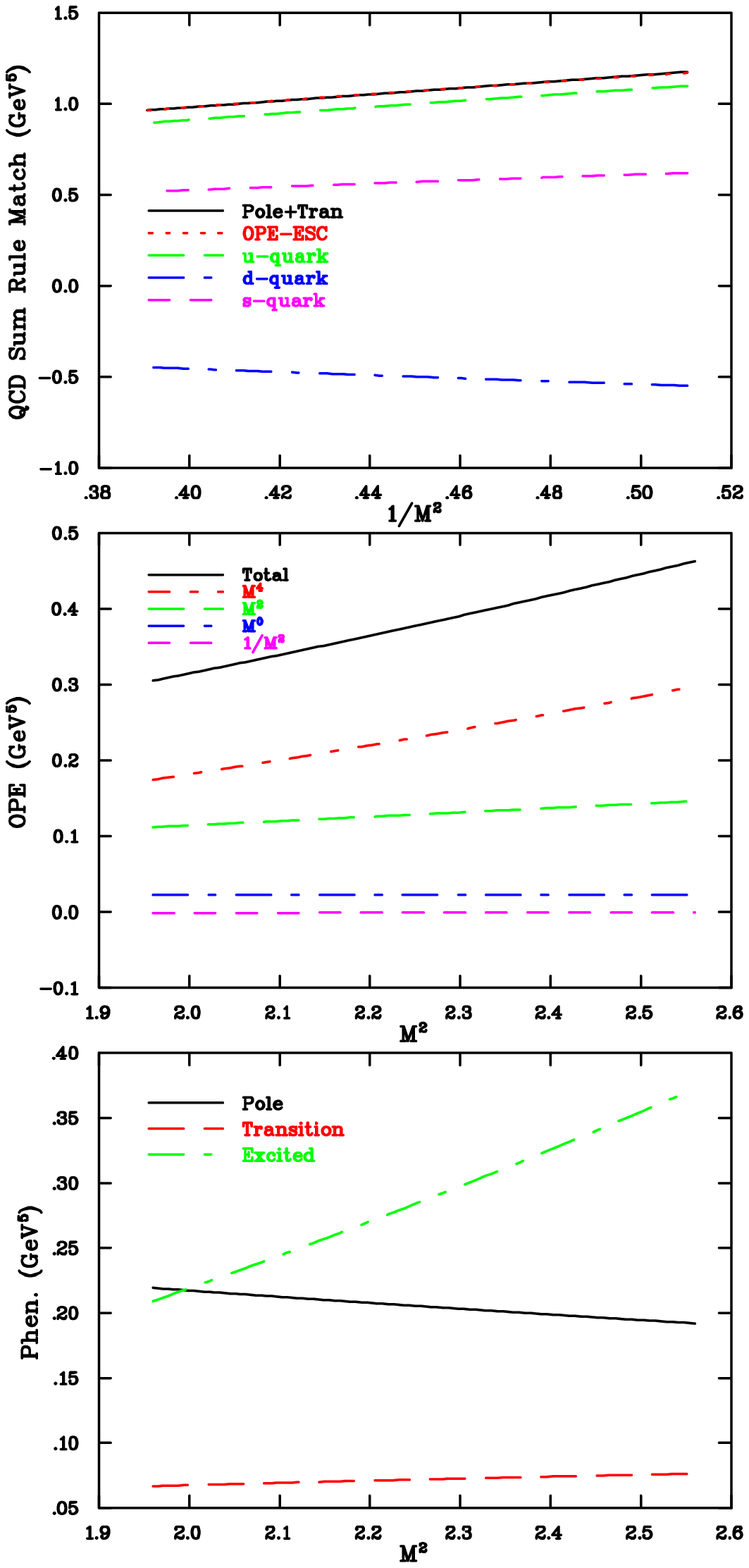, scale=0.45}
      \end{center}
  \end{minipage}
  \begin{minipage}[t]{.3\textwidth}
    \begin{center}
      \epsfig{file=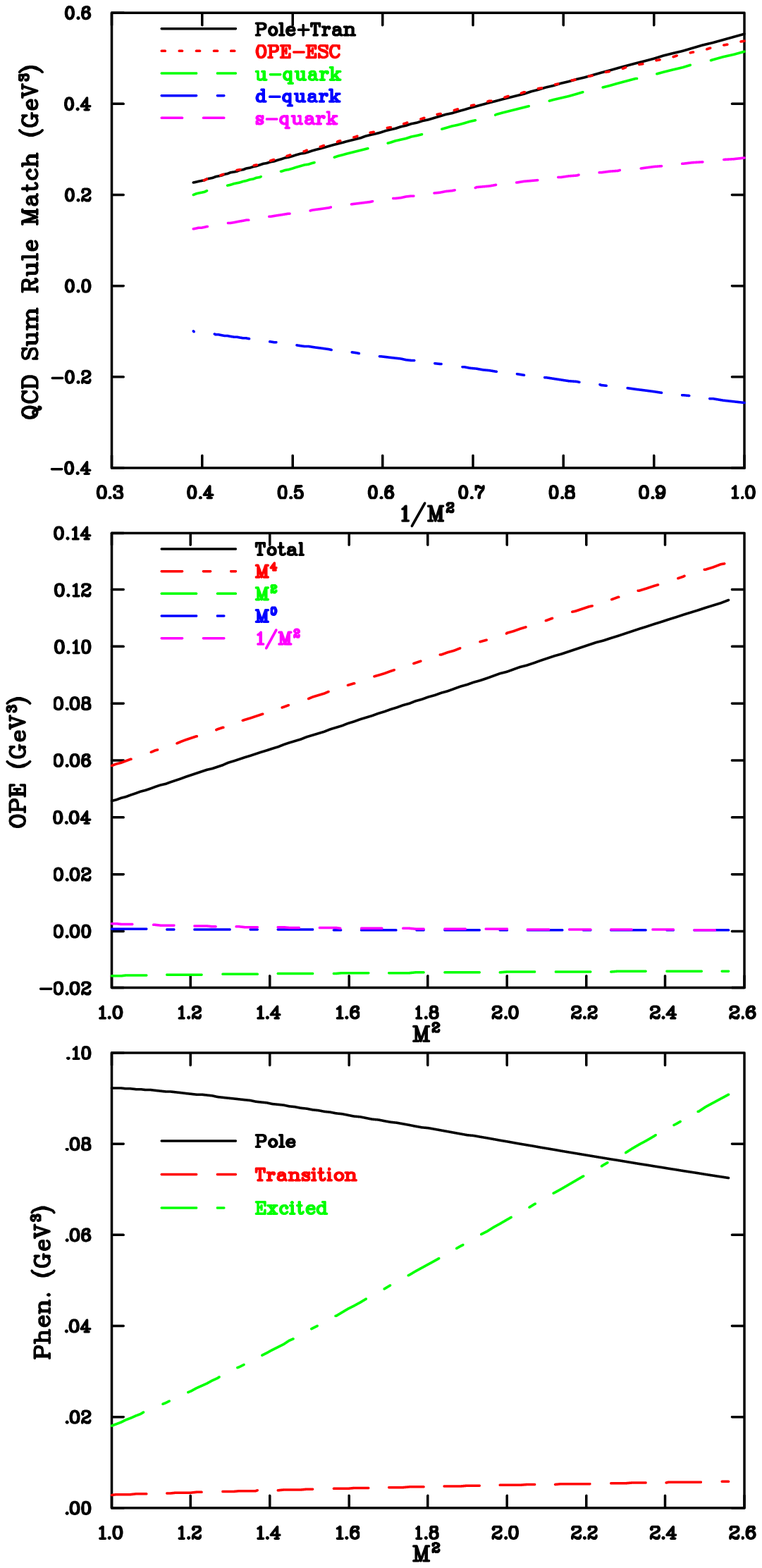, scale=0.45}
      \end{center}
  \end{minipage}
  \caption{Similar to Fig.~\protect\ref{nfig}, but for $\Sigma^0$.}
  \label{s0fig}
\end{figure*}

We have performed the same analysis for all the members and all
three structures. Fig.~\ref{nfig} shows the graphs for the neutron.
In this case, the slope is negative. Again, the sum rule at
$\mbox{WE}_1$ has excellent convergence properties. The
$\mbox{WO}_1$ sum rule has a good match, but the pole is less than
the excited-state. The $\mbox{WO}_2$ sum rule does not have a good
match. Fig.~\ref{s0fig} shows the case for the $\Sigma^0$, which has
all three quark contributions (u, d and s).

Overall, based on the quality of the match, the broadness of the Borel window and its
reach into the lower end, the size of the continuum contribution,
and the OPE convergence, we find that the sum rule
at $\mbox{WE}_1$ is the most reliable of the three sum rules.

\subsection{Some physics discussions}

Based on the results of our comprehensive analysis, we conclude that
the QCD sum rules at $\mbox{WE}_1$ in Eq.~(\ref{pwe1}) are the most reliable.
Here we discuss some physics implications of the results extracted
from them (Table~\ref{tabwe1}).
%
\begin{table}[hb]
\caption{A comparison of selected ratios of magnetic moments.}
\label{SU6}
 \begin{tabular*}{0.4\textwidth}{@{\extracolsep{\fill}}lccccc}
\hline\hline
              \vspace{-0.5mm}
              &Ratio  & Sum Rule    &  SU(6)    & Lattice & Exp.\\

                & & result   &  symmetry     &  results&\\
                \hline
&$n/p$   &-0.70(9)    &   $-2/3$      &-0.63(5)&-0.68\\
&$\Sigma^-/\Sigma^+$ &-0.50(7)   &     $-1/3$      &-0.37(3)& -0.47\\
&$\Xi^-/\Xi^0$   &0.56(6)      &  $1/2$       &0.58(5)&0.52\\
&$\Xi^0/\Lambda$   &2.05(56)      &  $2$       &2.4(5)&2.04\\
&$\Xi^-/\Lambda$   &1.14(32)      &  $1$       &1.0(2)&1.13\\
\hline\hline\\
\end{tabular*}
\end{table}
First we look at some ratios of magnetic moments
listed in Table~\ref{SU6}.
Here we compare our results with those from the SU(6) symmetry,
lattice calculations~\cite{Derek91} and experiment. From the
table, we see that the QCD sum rule results compare well against
other approaches and experiment. They agree a little better with experiment than the
lattice results. Furthermore, our QCD rum rule
results are an improvement over the ratios from previous QCD sum rule
calculations in Ref.~\cite{Chiu86}.

Next, we consider a few sum rules among magnetic moments that have been discussed
in literature. They reveal interesting quark dynamics in the baryons.
They are mostly based on SU(6)-symmetry considerations in the quark model.
We begin with the sum rule~\cite{Lipkin81}
\begin{equation}
\frac{3(p-\Sigma^+)}{\Xi^- -\Xi^0}=\frac{(p+3\Lambda)}{p}.
\label{SR1}
\end{equation}
It assumed `baryon independence' of quark moments: the independence
of which baryon the same quark is in, a concept first mentioned by
Franklin~\cite{Franklin69}. In other words, each quark is not
sensitive to the environment it resides in and quarks in different
spin states had the same effective moments. In this sum rule, it
also assumed that the s quark moment in the $\Lambda$ is the same as
in the $\Sigma$ and $\Xi$, even though the spin states are
different. This rum rule is violated the most (by a factor of 5)
using present values of magnetic moments. This large violation is
mostly due to the small difference in the denominator ${\Xi^-
-\Xi^0}$ which magnifies the apparent discrepancy. So this is not a
good way of testing baryon independence. Our determination of the
ratio of the left hand side over the right hand side (LHS/RHS) is
$7.4(1.0)$, compared to the experiment value of $5.7(4)$ and the
lattice calculation of $4.1(1.5)$. The error are added in quadrature
in forming these ratios.

One interesting sum rule~\cite{Franklin84} which is fairly
accurately satisfied is
\begin{equation}
p+n=3\Lambda+\frac{1}{2}(\Sigma^++\Sigma^-)-(\Xi^0+\Xi^-).
\label{SR2}
\end{equation}
It is derived using SU(3)-flavor symmetry
to characterize the $\Lambda$ wave function
Our determination of the LHS/RHS ratio is $1.24(79)$, compared to
the experimental values of $1.22(5)$ and the lattice calculation of
$1.45(27)$.

Another sum rule, first derived by Franklin~\cite{Franklin69},
\begin{equation}
p-n=\Sigma^+-\Sigma^-+\Xi^--\Xi^0,
\end{equation}
is another test of baryon independence of the quark
moments. The strange quarks approximately cancel, leaving only u and d quarks.
We give the ratio of $1.20(0.68)$, while the experiment
measurements gives $1.133(10)$ and lattice results is $1.07(8)$ for
this ratio.

Finally, the Sachs sum rule~\cite{Sachs81}
\begin{equation}
3(p+n)=\Sigma^+ - \Sigma^- - \Xi^- +\Xi^0
\end{equation}
is satisfied by the more general extension of SU(6) symmetry and is
another test of baryon independence of quark moments. This sum rule
is just the sum of the two separate sum rules proposed earlier by
Franklin~\cite{Franklin79}. Our result gives the ratio of $0.86(35)$
which is in agreement with the experiment measurements of
$0.881(11)$. Whereas it yields an opposite violation with a ratio of
$1.29(20)$ from lattice moments. This contradiction is possibly
because the Sachs sum rule may be sensitive to dynamics not included
in the lattice calculation.

\subsection{Individual quark contributions}

To gain a deeper understanding of the dynamics, it is useful to consider the
individual quark sector contributions to the magnetic moment. In our
approach, we can easily dial individual quark contributions to the
QCD sum rules. For example, to turn off all u-quark contributions,
we set the charge factor $c_u=0$. To turn off all s-quark
contributions, we set $c_s=0$, $m_s=0$, $f=1$, and $\phi=1$. We can
extract a number corresponding to each quark contribution from the
slope of Eq.~(\ref{ratio}) as a function of $1/M^2$. We call this
the raw individual quark contributions to the magnetic moments.
\begin{table*}[thb] %
\caption{Individual quark contributions and total magnetic moments
in unit of nuclear magnetons extracted from the QCD sum rules at
$\mbox{WE}_1$ (denoted by the subscript SR) are compared with those
from a lattice QCD calculation (denoted by the subscript
LAT)~\protect\cite{Derek91, Derek92}.} \label{quark}
 \begin{tabular*}{0.85\textwidth}{@{\extracolsep{\fill}}lcccccccc}
\hline\hline
                 &$\mu^u_{\texttt{SR}}$ & $\mu^u_{\texttt{LAT}}$
                 & $\mu^d_{\texttt{SR}}$  & $\mu^d_{\texttt{LAT}}$ & $\mu^s_{\texttt{SR}}$
                  & $\mu^s_{\texttt{LAT}}$  & $\mu^{Total}_{SR}$& $\mu^{Total}_{\texttt{LAT}}$  \\
                \hline
$p$ \hspace{1mm}  &2.46(25)   &2.59(24)   &0.38(4)    &0.15(9)    &0  &0  &2.82(26)   &2.79\\
$n$
\hspace{1mm}     &-0.42(9)   &-0.31(20)  &-1.53(13)  &-1.32(14)  &0  &0  &-1.97(15)  &-1.60(21)\\
$\Lambda$ \hspace{1mm} &0.18(7)    &0.09(6)    &-0.09(4)   &-0.03(2)    &-0.63(15)  &-0.54(7)   &-0.56(15)  &-0.50(7)\\
$\Sigma^+$
\hspace{1mm}    &2.06(25)   &2.33(29)   &0  &0  &0.15(2)    &0.08(5)    &2.31(25)   &2.37(18)\\
 $\Sigma^0$
     \hspace{1mm}&1.09(12)   &0.87(9)    &-0.54(6)   &-0.29(3)    &0.14(2)    &0.08(5)    &0.69(7)    &0.65(6)\\\
 $\Sigma^-$
 \hspace{1mm} &0  &0  &-1.25(10)  &-1.14(14)  &0.10(2) &0.08(5)    &-1.16(10)  &-1.08(10)\\
$\Xi^0$
\hspace{1mm}&0.02(2)    &-0.44(6)   &0  &0  &-1.08(7)   &-0.73(6)   &-1.15(5)   &-1.17(9)\\
 $\Xi^-$
\hspace{1mm} &0  &0  &0.03(2)    &0.22(3)    &-0.67(6)   &-0.73(6)   &-0.64(6)   &-0.51(7)\\
\hline\hline\\
\end{tabular*}
\end{table*}
%

Table~\ref{quark} gives the result of raw individual u, d and s
quark sector contributions to the magnetic moments from the QCD sum
rules at $\mbox{WE}_1$. It is compared with the lattice QCD result
in~\protect\cite{Derek91}. The lattice results were rescaled later
in \protect\cite{Derek92} and we use the rescaled results for the
comparison. Our results agree with the lattice results reasonably
well.  The biggest discrepancy is that our light quark moments in
$\Xi^0$ and $\Xi^-$ are small. In $\Sigma^0$, and $\Lambda$ the
total of u and d quark moments agree very well with lattice data for
light quark moments. Note that only a combined number for light
quark ($\mu_l=\mu_u+\mu_d$) was given on the lattice~\cite{Derek91}.
To compare with our separated u and d quark numbers, we break up the
lattice number by using the relation $\mu_u=-2\mu_d$.

In the simple quark model~\cite{Perkins}, the magnetic moment of the
proton is given by $p=\frac{4}{3}\mu_u-\frac{1}{3}\mu_d$. In the
SU(2) limit, there is the relation $\mu_u=-2\mu_d$. Our results for
$\Sigma^0$ and $\Lambda$, the two $uds$ structure baryons, agree
with this very well. This relation also suggests that the ratio of
the u quark and d quark contributions in proton is
$\frac{4}{3}\mu_u/({-\frac{3}{4}\mu_d})=8$. Our QCD sum rule ratio
is $10.6$ for $\beta=-1.2$, which agrees with the lattice ratio of
$10.3(7)$ at $\kappa_1$. For neutron the quark model ratio is
$\frac{4}{3}\mu_d/({-\frac{1}{3}\mu_u})=2$, and our result is
$3.64(40)$ for this ratio, which differs from the SU(6) spin-flavor
symmetry prediction. The lattice ratio is $2.6$. The bigger ratio
represents an enhancement of the doubly represented quark
contribution. These results suggest that although the total magnetic
moments ratios agree with SU(6) ratios (see Table~\ref{SU6}), the
underlying quark dynamics are really quite different from different
individual quark contributions.

To include the possible nonstatic effect, Franklin~\cite{Franklin84}
proposed a generalization of SU(6) results so that nonstatic
components are the same for each octet baryon:
\begin{equation}
\begin{array}{*{20}l}
p=\frac{4}{3}\mu_u-\frac{1}{3}\mu'_d,\\
n=\frac{4}{3}\mu_d-\frac{1}{3}\mu'_u,\\
\Sigma^+=\frac{4}{3}\mu_u-\frac{1}{3}\mu'_s,\\
\Sigma^-=\frac{4}{3}\mu_d-\frac{1}{3}\mu'_s,\\
\Xi^0=\frac{4}{3}\mu_s-\frac{1}{3}\mu'_u,\\
\Xi^-=\frac{4}{3}\mu_s-\frac{1}{3}\mu'_d,\\
\Sigma^0=\frac{2}{3}\mu_u+\frac{2}{3}\mu_d-\frac{1}{3}\mu'_s,\\
\Lambda=\mu''_s\\
\end{array}\label{quarkmodel}
\end{equation}
where quark symbols refer to quark moment contributions including
nonstatic effect. The magnetic moment contribution of the unlike
quark in the baryon is primed.

In order to compare with the effective quark moments defined in the
quark model, we convert our raw quark moments in Table~{quark} in a
similar fashion. Take the proton for example, we define effective
quark moments $\mu_u$ and $\mu'_d$ from QCD sum rules by
$u_{SR}={4\over 3} \mu_u$, $d_{SR}={1\over 3} \mu'_d$. One can
define effective moments for quarks in other baryons in a similar
manner.

In the proton, $\mu_u$ is $1.84(18)$ while in $\Sigma^+$  $1.54(18)$
and in $\Sigma^0$ is $1.63(18)$. The magnetic moments list here are all in
unit of $\mu_N$. In the neutron, $\mu'_u$ is $1.25(27)$, which is smaller
than $\mu_u$. This is an example that
effective quark magnetic moment is sensitive to the environment the quark resides in.
We find the  following relation from our results
\begin{equation}
\mu^p_u < \mu^{\Sigma^0}_u< \mu^{\Sigma^+}_u.
\end{equation}
For $\mu_d$,  in $n$ it is $-1.15(11)$ while for $\Sigma^0$ is
$-0.82(8)$ and for $\Sigma^-$ is $-0.93(8)$. From $p$ we can get
$\mu'_d=-1.15(12)$. It has a similar relation for the absolute value
\begin{equation}
\mu^n_d < \mu^{\Sigma^0}_d< \mu^{\Sigma^-}_d.
\end{equation}

For $\mu_s$, it is $-0.81(6)$ in $\Xi^0$ and $-0.50(5)$ in $\Xi^-$.
For $\mu'_s$, it is $-0.45(5)$, $-0.42(5)$
and $-0.30(5)$ in $\Sigma^+$, $\Sigma^0$ and $\Sigma^-$,
respectively. From $\Lambda$ we can get $\mu''_s$ about $-0.63(15)$.

For these individual quark effective moments, we notice that in
general we have the following relation

\begin{equation}
\mu_s < \mu_d< \mu_u.
\end{equation}
It is expected because of the quark mass effects, which is analogous
to those seen in the electric properties.

Another way of looking at the individual effective quark moments is
by expressing them in terms of baryon magnetic
moments using Eq.(\ref{quarkmodel}) and isospin symmetry.
For example, the d-quark effective moment can be expressed as
\begin{equation}
\mu_d = -\frac{1}{4}(2p+n)=\frac{1}{4}(\Sigma^--\Sigma^+).
\end{equation}
Our result indicates $\mu_d=-0.92(14)<-0.86(7)$ using the baryon
moments in Table~\ref{quark} or in Table~\ref{tabwe1}. It agrees
well with the experimental moments $\mu_d=-0.918<-0.894(7)$ and the
lattice result of $\mu_d=-1.00(5)<-0.86(6)$.

Similarly the $d'$ quark effective moment is
\begin{equation}
\mu'_d = p+2n=\Xi^0-\Xi^-.
\end{equation}
We have $\mu'_d=-1.12(39)<-0.51(8)$. It also agree very well with
the experimental moments $\mu'_d=-1.003<-0.578(26)$ although the two
sides of this sum don't agree well.

The strange quark can be isolated as
\begin{equation}
\begin{array}{*{20}l}
\mu_s =\frac{1}{4}(\Xi^0+2\Xi^-),\\
\mu'_s=-\Sigma^+-2\Sigma^-,\\
\mu''_s=\Lambda-\delta.
\end{array}
\end{equation}
Our results are compared with experiment and lattice calculation in Table~\ref{squark}.
\begin{table}[thb] %
 \caption{The result for $s$ $s'$ and $s''$ quark effective
moments from QCD sum rule, experiment and lattice calculation.}
\label{squark}
\begin{tabular}{lccccc}
\hline\hline
                 &$\mu_s$&&$\mu''_s$&&$\mu'_s$ \\
                \hline
SR \hspace{1mm}  &-0.61(3)  &$<$ &-0.60(15)   &$<$ &0.01(32)\\
Exp.
\hspace{1mm}     &-0.651(12)  &$<$ &-0.57(4)   &$<$& -0.107(36)\\
LAT.
\hspace{1mm}&-0.55(4)  &$\approx $ &-0.54(4)   &$<$ &-0.24(17)\\
\hline\hline\\
\vspace{-8mm}
\end{tabular}
\end{table}
Not only the results agree with the experiment and lattice
calculation, they also roughly agree with the strange quark moments
from individual quark moments.

Now, we can look at the quark moment difference such as
$\mu'_s-\mu'_d$ and $\mu_s-\mu_d$. According to simple quark model,
the difference should be the same and approximately as $0.36$. In
fact, from
\begin{equation}
\mu'_s-\mu'_d = 3(p-\Sigma^+),
\label{udprime}
\end{equation}
\begin{equation}
\mu_s-\mu_d = \frac{3}{4}(\Xi^0-n), \label{ud}
\end{equation}
QCD sum rules give $1.53(8)$ and $0.61(12)$. The experiment gives
$1.12(7)$ and $0.495(11)$, while lattice calculation gives $1.3(5)$ and
$0.32(17)$ respectively. The large result of
$\mu'_s-\mu'_d$ is difficult to reconcile with any simple model. And
Eq.~\ref{udprime} is a poor way to measure the difference of s and d
quark contribution.
%
\subsection{Correlations}
Our Monte-Carlo analysis affords the opportunity to study the
correlations between any two parameters since the entire QCD input
phase space is mapped into the phenomenological output space. This
correlation can be explored by a scatter plot of the two parameters
of interest.

\begin{figure*}[tbh]
    \begin{center}
      \epsfig{file=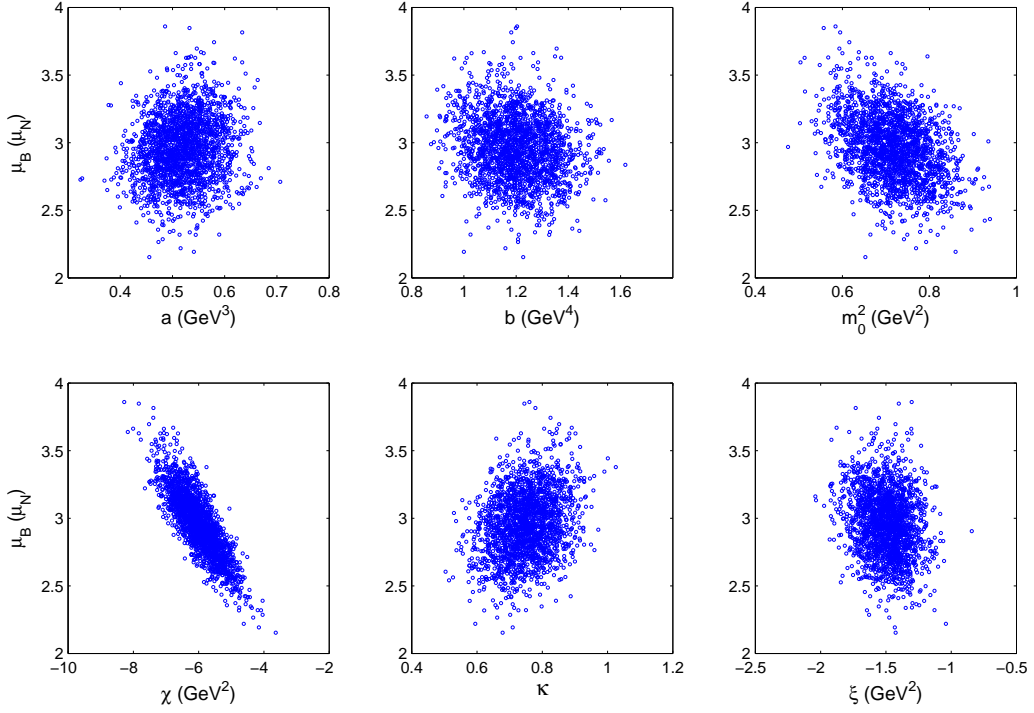, scale=0.60}
      \end{center}
\vspace{-1cm}
  \caption{Scatter plots showing correlations between the magnetic moment
and the QCD parameters for the proton at structure $\mbox{WE}_1$.
They are obtained from 2000 Monte-Carlo samples with 10\% uncertainty on all the
QCD parameters.}
  \label{scapwe1}
\end{figure*}

Fig.~\ref{scapwe1} shows the scatter plots for the proton magnetic moment
at structure $\mbox{WE}_1$.
We see that the magnetic moment has a strong correlation with the vacuum
susceptibility $\chi$. It is a negative correlation meaning larger $\chi$ (
in absolute terms since $\chi$ is negative) leads to smaller $\mu_B$.
A slight negative correlation with the mixed condensate and a slight
positive correlation
with another vacuum susceptibility $\kappa$ are also observed.
Precise determination of the QCD parameters,
especially for those that have strong correlations to the output parameters,
is crucial for keeping the uncertainties in the spectral parameters under control.

\begin{figure*}[tbh]
    \begin{center}
      \epsfig{file=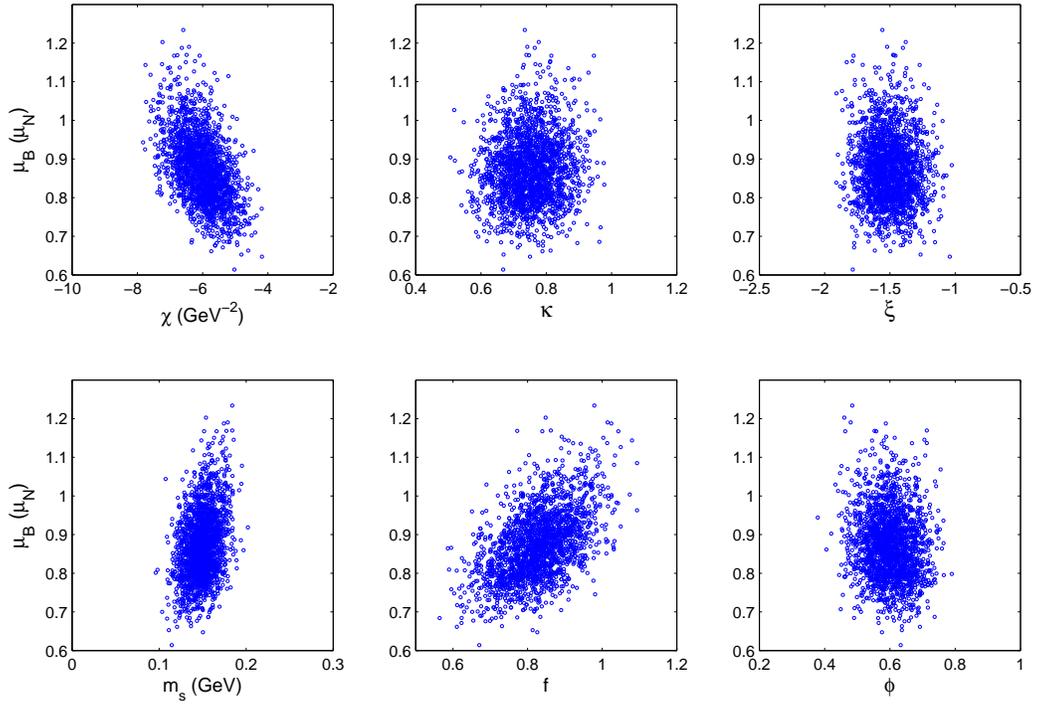, scale=0.60}
      \end{center}
\vspace{-1cm}
  \caption{Similar to Fig.~\protect\ref{scapwe1}, but for the $\Sigma^0$
and a different set of QCD parameters.}
  \label{scas0we1}
\end{figure*}

Fig.~\ref{scas0we1} shows a similar plot for the $\Sigma^0$ at structure $\mbox{WE}_1$.
Here we focus on the three vacuum susceptibilities and the three parameters that
define the strange quark ($m_s$, $f$ and $\phi$).
The correlations with the other condensates are similar to the proton and are not shown.
A negative correlation with $\chi$ exists, but not as strong as that for the proton.
A slight positive correlation with $m_s$ and $f$ is also observed.

\section{Conclusion}
\label{con}
We have carried out a comprehensive study of the magnetic moment of octet baryons
using the method of QCD sum rules.
We derived a new, complete set of QCD sum rules using generalized interpolating fields
and examined them by a Monte-Carlo analysis.
Here is a summary of our findings.

We proposed a new way of determining the magnetic moments from the
slope of straight lines. We find this method more robust than from
the normalization (intercept) or from looking for `flatness' as a
function of Borel mass. The linearity displayed from the OPE side
matches almost perfectly with the phenomenological side in most
cases. The method also demonstrates clearly that the transition
terms caused by the external field in the intermediate terms cannot
be ignored. They are needed to make the two sides of a QCD sum rule
match.

Out of the three independent structures, we find that the sum rules from
the $\mbox{WE}_1$ structure are the most reliable based on OPE convergence and
ground-state pole dominance.
The QCD sum rules from this structure are in Eq.~(\ref{we1});
its prediction are found in Table~\ref{tabwe1}, and
convergence properties are displayed in Fig.~\ref{pwe1}.
They should be considered as the best results in this work.
The extracted magnetic moments are in good agreement with experiment.
These results are used to shed light on a variety of magnetic moment
sum rules based on SU(6) spin-flavor symmetries in the quark model,
along with experiment and lattice QCD.
Reasonable results from the other two structures ($\mbox{WO}_1$ and $\mbox{WO}_2$)
are obtained for the first time, but they are less reliable due to poor convergence
properties.

Our Monte-Carlo analysis revealed that there is an uncertainty
on the level of 10\% in the magnetic moments if we assign 10\% uncertainty
in the QCD input parameters. It goes up to about 30\% if we adopt the
conservative assignments that have a wide range of uncertainties in Ref.~\cite{Derek96}.
The Monte-Carlo analysis also revealed some correlations between the input and output
parameters. The most sensitive is the vacuum susceptibility $\chi$. So a better determination
of this parameter can help improve the accuracy on the magnetic moments and other
quantities computed from the same method.

We also isolated the individual quark contributions to the
magnetic moments. These contributions provide insight into the rich dynamics in the baryons.
By comparing them with the simple quark model and lattice QCD results,
we reveal the effects of SU(3)-flavor symmetry breakings in the strange quark,
the environment sensitivity of quarks in different baryons.

Taken together, this work can be considered an updated and improved determination of the
magnetic moments of octet baryons,
bringing it to the same level of sophistication as the decuplet baryons~\cite{Lee98b}.
One possible extension along this line is a calculation of the N to $\Delta$
electromagnetic transition amplitudes which to our knowledge
have not been studied in this method.

\begin{acknowledgments}
This work is supported in part by U.S. Department of Energy under grant
DE-FG02-95ER-40907.
\end{acknowledgments}


%
%

%
%
%
\end{document}